\documentclass[twocolumn,twocolappendix]{aastex631}

\usepackage{amsmath}
\usepackage{wasysym}
\usepackage{caption}
\usepackage{placeins}

\newcommand{\rev}[1]{#1}

\newcommand\extrafootertext[1]{%
    \bgroup
    \renewcommand\thefootnote{\fnsymbol{footnote}}%
    \renewcommand\thempfootnote{\fnsymbol{mpfootnote}}%
    \footnotetext[0]{#1}%
    \egroup
}

\newcommand{\ChainConsumer}{\href{https://samreay.github.io/ChainConsumer/\#}{\texttt{ChainConsumer}}}
\newcommand{\corner}{\href{https://corner.readthedocs.io/en/latest/index.html\#}{\texttt{corner}}}
\newcommand{\DRAC}{\href{https://alliancecan.ca/en}{Digital Research Alliance of Canada (DRAC)}}
\newcommand{\KeplerSpaceTelescope}{\href{https://astrobiology.nasa.gov/missions/kepler/}{0.95m \ik Satellite Mission}}
\newcommand{\Matplotlib}{\href{https://matplotlib.org/}{\texttt{Matplotlib}}}
\newcommand{\NASAADS}{\href{https://ui.adsabs.harvard.edu/}{Astrophysics Data System (NASA ADS)}}
\newcommand{\NASAExoplanetArchive}{\href{https://exoplanetarchive.ipac.caltech.edu/index.html}{NASA Exoplanet Archive}}
\newcommand{\Numba}{\href{https://numba.readthedocs.io/en/stable/index.html}{\texttt{Numba}}}
\newcommand{\NumPy}{\href{https://numpy.org/}{\texttt{NumPy}}}
\newcommand{\pandas}{\href{https://pandas.pydata.org/}{\texttt{pandas}}}
\newcommand{\Python}{\href{https://www.python.org/}{\texttt{Python}}}
\newcommand{\quantile}{\href{https://corner.readthedocs.io/en/latest/api/\#corner.quantile}{\texttt{quantile}}}
\newcommand{\ReactiveNestedSampler}{\href{https://johannesbuchner.github.io/UltraNest/ultranest.html?highlight=reactivenestedsampler\#ultranest.integrator.ReactiveNestedSampler}{\texttt{ReactiveNestedSampler}}}
\newcommand{\RegionSliceSampler}{\href{https://johannesbuchner.github.io/UltraNest/ultranest.html?highlight=regionslicesampler\#ultranest.stepsampler.RegionSliceSampler}{\texttt{RegionSliceSampler}}}
\newcommand{\SciPy}{\href{https://scipy.org/}{\texttt{SciPy}}}

\newcommand{\transitfitfive}{\href{https://zenodo.org/record/60297}{\texttt{transitfit5}}}
\newcommand{\UltraNest}{\href{https://johannesbuchner.github.io/UltraNest/}{\texttt{UltraNest}}}
\newcommand{\UAT}{\href{https://astrothesaurus.org/}{Unified Astronomy Thesaurus}}

\newcommand{\data}{\ensuremath{\mathcal{D}}}
\newcommand{\evidence}{\hyperlink{def:evidence}{\ensuremath{\mathcal{Z_{D}}}}}
\newcommand{\likelihood}{\hyperlink{def:likelihood}{\ensuremath{\mathcal{L_{D}} \left( \theta \right)}}}
\newcommand{\parameters}{\ensuremath{\theta}}
\newcommand{\posteriors}{\hyperlink{def:posteriors}{\ensuremath{\mathcal{P_{D}} \left( \theta \right)}}}
\newcommand{\priors}{\hyperlink{def:priors}{\ensuremath{\pi \left( \theta \right)}}}

\newcommand{\logB}{\hyperlink{def:logB}{\ensuremath{\mathcal{B}}}}

\newcommand{\days}{\ensuremath{\rm days}}
\newcommand{\gcmthree}{\ensuremath{\rm g \ cm^{- 3}}}
\newcommand{\hours}{\ensuremath{\rm hours}}
\newcommand{\Kelvin}{\ensuremath{\rm K}}

\renewcommand{\Earth}{\ensuremath{\boldsymbol{\rm \oplus}}}
\renewcommand{\Sun}{\ensuremath{\boldsymbol{\rm \odot}}}
\renewcommand{\Venus}{\ensuremath{\boldsymbol{\rm \venus}}}

\newcommand{\PeriodEarth}{\ensuremath{\rm P_{\Earth}}}
\newcommand{\REarth}{\ensuremath{\rm R_{\Earth}}}
\newcommand{\SzeroEarth}{\ensuremath{\rm S_{\rm 0 , \Earth}}}

\newcommand{\LSun}{\ensuremath{\rm L_{\Sun}}}
\newcommand{\MSun}{\ensuremath{\rm M_{\Sun}}}

\newcommand{\RSun}{\ensuremath{\rm R_{\Sun}}}

\newcommand{\rhostar}{\hyperlink{def:rhostar}{\ensuremath{\rho_{\rm \star}}}}
\newcommand{\qone}{\hyperlink{def:qone}{\ensuremath{q_{\rm 1}}}}
\newcommand{\qtwo}{\hyperlink{def:qtwo}{\ensuremath{q_{\rm 2}}}}
\newcommand{\Tzero}{\hyperlink{def:Tzero}{\ensuremath{T_{\rm 0}}}}
\newcommand{\Period}{\hyperlink{def:Period}{\ensuremath{P}}}
\newcommand{\bimpact}{\hyperlink{def:bimpact}{\ensuremath{b}}}
\newcommand{\RpRstar}{\hyperlink{def:RpRstar}{\ensuremath{\Rp{} / \Rstar{}}}}
\newcommand{\Fzero}{\hyperlink{def:Fzero}{\ensuremath{F_{\rm 0}}}}
\newcommand{\sigmaw}{\hyperlink{def:sigmaw}{\ensuremath{\sigma_{\rm w}}}}
\newcommand{\sigmac}{\hyperlink{def:sigmac}{\ensuremath{\sigma_{\rm c}}}}
\newcommand{\lc}{\hyperlink{def:lc}{\ensuremath{l_{c}}}}

\newcommand{\ecosw}{\hyperlink{def:ecosw}{\ensuremath{\sqrt{\eccentricity{}} \cos \left( \wperi{} \right)}}}
\newcommand{\esinw}{\hyperlink{def:esinw}{\ensuremath{\sqrt{\eccentricity{}} \sin \left( \wperi{} \right)}}}

\newcommand{\aRstar}{\hyperlink{def:aRstar}{\ensuremath{a / \Rstar{}}}}
\newcommand{\Deltat}{\ensuremath{\Delta t}}
\newcommand{\etaEarth}{\hyperlink{def:etaEarth}{\ensuremath{\eta_{\Earth}}}}
\newcommand{\etaVenus}{\hyperlink{def:etaVenus}{\ensuremath{\eta_{\Venus}}}}
\newcommand{\FA}{\hyperlink{def:FA}{\ensuremath{\rm FA}}}
\newcommand{\FP}{\hyperlink{def:FP}{\ensuremath{\rm FP}}}
\newcommand{\KOI}{\hyperlink{def:KOI}{\ensuremath{\rm KOI}}}
\newcommand{\EB}{\ensuremath{\rm EB}}
\newcommand{\eccentricity}{\hyperlink{def:eccentricity}{\ensuremath{e}}}
\newcommand{\eccentricitymin}{\ensuremath{e_{\rm min}}}
\newcommand{\GP}{\hyperlink{def:GP}{\ensuremath{\rm GP}}}
\newcommand{\PC}{\hyperlink{def:PC}{\ensuremath{\rm PC}}}
\newcommand{\MAP}{\hyperlink{def:MAP}{\ensuremath{\rm MAP}}}
\newcommand{\MES}{\hyperlink{def:MES}{\ensuremath{\rm MES}}}
\newcommand{\Mp}{\ensuremath{M_{\rm p}}}
\newcommand{\Rp}{\hyperlink{def:Rp}{\ensuremath{R_{\rm p}}}}
\newcommand{\Rpprime}{\hyperlink{def:Rp}{\ensuremath{R_{\rm p}^{\prime}}}}
\newcommand{\sigmad}{\hyperlink{def:sigmad}{\ensuremath{\sigma_{\rm d}}}}
\newcommand{\sigmap}{\hyperlink{def:sigmap}{\ensuremath{\sigma_{\rm p}}}}
\newcommand{\Lstar}{\ensuremath{L_{\rm \star}}}
\newcommand{\Mstar}{\ensuremath{M_{\rm \star}}}

\newcommand{\Rstar}{\hyperlink{def:Rstar}{\ensuremath{R_{\rm \star}}}}
\newcommand{\Szero}{\hyperlink{def:Szero}{\ensuremath{S_{\rm 0}}}}
\newcommand{\SBI}{\ensuremath{\rm SBI}}
\newcommand{\SNR}{\hyperlink{def:SNR}{\ensuremath{\rm S/N}}}
\newcommand{\SNRprime}{\hyperlink{def:SNR}{\ensuremath{\rm S/N^{\prime}}}}

\newcommand{\TCE}{\hyperlink{def:TCE}{\ensuremath{\rm TCE}}}
\newcommand{\Teff}{\hyperlink{def:Teff}{\ensuremath{T_{\rm eff}}}}
\newcommand{\TGP}{\hyperlink{def:TGP}{\ensuremath{\rm TGP}}}
\newcommand{\tT}{\hyperlink{def:tT}{\ensuremath{t_{\rm T}}}}
\newcommand{\tTc}{\hyperlink{def:tTc}{\ensuremath{t_{\rm T , c}}}}
\newcommand{\wperi}{\hyperlink{def:wperi}{\ensuremath{\omega}}}

\newcommand{\ikt}{\textit{Kepler}}
\newcommand{\ik}{\textit{Kepler~}}
\newcommand{\Ktwot}{\textit{K2}}
\newcommand{\tesst}{\textit{TESS}}

\received{2023 September 9}
\revised{2023 November 20}
\accepted{2023 November 25}
\published{2024 January 19}

\submitjournal{\aj}

\shorttitle{Bayesian Assessment of Kepler’s Exoplanet Candidates}
\shortauthors{Matesic et al.}

\begin{document}

\title{\large Gaussian processes and Nested Sampling Applied to Kepler's Small Long-Period Exoplanet Candidates}

\correspondingauthor{Michael R. B. Matesic}
\email{michael.matesic@umontreal.ca}

\author[0000-0002-1119-7473]{Michael R. B. Matesic}
\affiliation{D{\'e}partement de Physique, Trottier Institute for Research on Exoplanets, Ciela Institute for Computation \& Astrophysical Data Analysis, Universit{\'e} de Montr{\'e}al, 1375 Th{\'e}r{\`e}se-Lavoie-Roux Av., Montr{\'e}al, QC H2V 0B3, Canada}
\affiliation{Department of Physics \& Astronomy, Bishop's University, 2600 Rue College, Sherbrooke, QC J1M 1Z7, Canada}

\author[0000-0002-5904-1865]{Jason F. Rowe}
\affiliation{Department of Physics \& Astronomy, Bishop's University, 2600 Rue College, Sherbrooke, QC J1M 1Z7, Canada}

\author[0000-0002-4881-3620]{John~H.~Livingston}
\affiliation{Astrobiology Center, 2-21-1 Osawa, Mitaka, Tokyo 181-8588, Japan}
\affiliation{National Astronomical Observatory of Japan, 2-21-1 Osawa, Mitaka, Tokyo 181-8588, Japan}
\affiliation{Department of Astronomical Science, School of Physical Sciences, The Graduate University for Advanced Studies (SOKENDAI), 2-21-1, Osawa, Mitaka, Tokyo, 181-8588, Japan}

\author[0000-0001-6263-4437]{Shishir Dholakia}
\affiliation{Centre for Astrophysics, University of Southern Queensland, 487-535 West St, Darling Heights, QLD 4350, Australia}

\author[0000-0002-6227-7510]{Daniel Jontof-Hutter}
\affiliation{Department of Physics, University of the Pacific, 3601 Pacific Avenue, Stockton, CA 95211, USA}

\author[0000-0001-6513-1659]{Jack J. Lissauer}
\affiliation{Space Science \& Astrobiology Division, MS 245-3, NASA Ames Research Center, Moffett Field, CA 94035, USA}

\begin{abstract}
    There are more than 5000 confirmed and validated planets beyond the Solar System to date, more than half of which were discovered by NASA's \ik mission. The catalog of \ikt's exoplanet candidates has only been extensively analyzed under the assumption of white noise (i.i.d. Gaussian), which breaks down on timescales longer than a day due to correlated noise (point-to-point correlation) from stellar variability and instrumental effects. Statistical validation of candidate transit events becomes increasingly difficult when they are contaminated by this form of correlated noise, especially in the low signal-to-noise (\SNR{}) regimes occupied by Earth--Sun and Venus--Sun analogs. To diagnose small long-period, low-\SNR{} putative transit signatures with few (roughly 3~--~9) observed transit-like events (e.g., Earth--Sun analogs), we model \ikt’s photometric data as noise, treated as a Gaussian process, with and without the inclusion of a transit model. Nested sampling algorithms from the \texttt{Python} \texttt{UltraNest} package recover model evidences and \textit{maximum a posteriori} parameter sets, allowing us to disposition transit signatures as either Planet-Candidates or False-Alarms within a Bayesian framework.
\end{abstract}


\section[Introduction]{Introduction} \label{sec:Introduction}
        
    The NASA \ik space telescope \citep{2010_02_Borucki,2010_04_Koch,2016_03_Borucki} launched in 2009 and observed $\sim 200,000$ stars within its primary field of view over the course of roughly 4 yr. With instrumental error budgets capable of detecting Earth-sized planets in year-long orbits around Sun-like stars, \ik aimed to directly measure the occurrence rate of such objects, otherwise known as eta-Earth \citep[\etaEarth{};][]{2010_02_Borucki}. Although the fulfillment of this objective was impeded by greater noise contamination from both stellar and instrumental effects than initially anticipated \citep{2011_11_Gilliland,2015_10_Gilliland}, significant progress has still been made. To aid in this effort, our work debuts a novel Bayesian framework employing nested sampling \citep{2004_11_Skilling,2006_12_Skilling} alongside simultaneous correlated noise modelling with Gaussian processes \citep[\GP{}s;][]{1999_06_Stein,2006_Rasmussen} to more accurately conduct Planet-Candidate (\PC{})-False-Alarm (\FA{}) dispositioning and characterization. As an aside, this study distinguishes \FA{}s---being instrumental or astrophysical variability which mimic transit events---from astrophysical False-Positives (\FP{}s)---being transit-like events produced by eclipsing binary stars (\EB{}s) and blends.

    Currently, no potential Earth--Sun or Venus--Sun analog system from the \ik sample has been shown to be reliable. Moreover, the occurrence rates for planets with $0.5 < \Rp{} < 1.75 \ \REarth{}$ and $64 < \Period{} < 500 \ \days{}$ as shown in Figure~2 of \citet{2019_09_Hsu} are either upper bounds or detections with statistical significance less than 2 standard deviations, so extrapolation to regions of parameter space with fewer candidates would incur large statistical uncertainties. Thus, the estimate of \etaEarth{} (and \etaVenus{}) can be improved via more robust reliability estimates not only in the Earth--Sun and Venus--Sun analog bins but also in adjacent bins containing few verified planets. \FA{}s, not astrophysical \FP{}s such as \EB{}s, become the primary issue for Kepler Object of Interest \citep[\KOI{};][]{2018_04_Thompson} discrimination in these regions \citep[see Figure~6 of][]{2018_04_Thompson}. Note that published \KOI{} catalogs do not distinguish between \FA{}s and \FP{}s, dispositioning both classes of objects as \FP{}s, because their purpose is to distinguish planet candidates from non-candidates.

    Having undergone thorough preconditioning via the Presearch Data Conditioning \citep[PDC;][]{2010_07_Twicken,2012_09_Stumpe,2012_09_Smith} module of the Kepler Science Operations Center (SOC) Science Processing Pipeline \citep{2010_04_Jenkins} in an attempt to mitigate instrumental trends common among all stars on the detector, the data products of \KOI{}s should ideally only contain intrinsic stellar variability (granulation, spots, flares, oscillations, etc.) and transiting exoplanet/eclipsing stellar binary signatures; however, instrumental systematics (sudden pixel sensitivity dropout, rolling band, bad pixels, cosmic rays, etc.)---which impact light curves non-uniformly---can also persist \citep{2010_04_Caldwell,2011_11_Gilliland,2014_06_Clarke,2015_10_Gilliland,2016_04_Van_Cleve,2019_06_Kawahara}. 

    As previous studies, such as Data Release 25 \citep[DR25;][]{2016_12_Twicken,2017_04_Mathur,2018_04_Thompson}, do not model the transit event and correlated noise simultaneously \textit{or} compute the individual reliability for any single target, their results are left susceptible to \FA{} misidentification \citep{2015_06_Foreman_Mackey}; instead, interpolation is performed across orbital period and Multiple Event Statistic (\MES{}) using population-level injection results \citep{2017_06_Christiansen,2020_06_Bryson}. \rev{Another example, \citet{2019_08_Caceres}, statistically classifies Earth-sized \ik \PC{}s in the presence of correlated noise. However, their approach is frequentist, does not reveal any long-period \PC{}s, and does not robustly estimate transit parameters.} By analyzing individual light curves on a per-target basis, our work better safeguards against \FA{}s while also improving the accuracy and robustness of \PC{} characterization in comparison to previous population-level approaches. 
    
    Accordingly, the data of any given \KOI{} can be interpreted as having originated from a transiting \PC{} with some noise contamination or as a purely noise \FA{}. To assess the probability that small long-period low-signal-to-noise ratio (\SNR{}) patterns of photometric dips with few (roughly 3~--~9) observed transit-like events (i.e., the regime that includes Earth--Sun and Venus--Sun analogs) are of astrophysical origin (i.e., represent true \PC{}s or background/hierarchical \EB{}s which induce transit-like dips), we model \ikt’s photometric data as noise, treated as a \GP{}, with and without the inclusion of a transit model. These are hereby denoted as the transit plus Gaussian process (\TGP{}) and \GP{} models, representing \PC{} and \FA{} interpretations, respectively; model parameters are described in \autoref{tab:Model Parameter Descriptions}. Here, two qualitatively different models are being compared: one with a pattern of transit-shaped dips (\TGP{}) and the other without (\GP{}). The former wields more degrees of freedom and accordingly will fit the data more closely, but we must ask whether these additional parameters are justified. To provide a principled answer, we employ Bayesian model comparison.
    
    Rooted in Bayes's theorem, nested sampling algorithms from the \Python{} \citep{1995_01_a_Van_Rossum,1995_01_b_Van_Rossum,1995_01_c_Van_Rossum,1995_01_d_Van_Rossum,1996_05_Dubois,2007_01_Oliphant} \UltraNest{} \citep{2016_01_Buchner,2019_11_Buchner,2021_04_Buchner} package recover \textit{maximum a posteriori} (\MAP{}) parameter sets and evidences of each model, allowing for transit signatures to be dispositioned in terms of \PC{} and \FA{} probabilities within a Bayesian framework. It is important to clarify that this work does not attempt to qualify \KOI{}s beyond \PC{} or \FA{} status; this is in sharp contrast to \ik planet catalogs, which disposition \FA{}s together with \FP{}s. 
    
    The simultaneous modeling of correlated noise additionally provides more robust constraints on transit model parameters. Thus, the analysis that we present herein also improves the characterization of \PC{}s, most significantly in terms of their radii. 

    \begin{figure*}[htb!]
        \centering
        \includegraphics[width = \hsize]{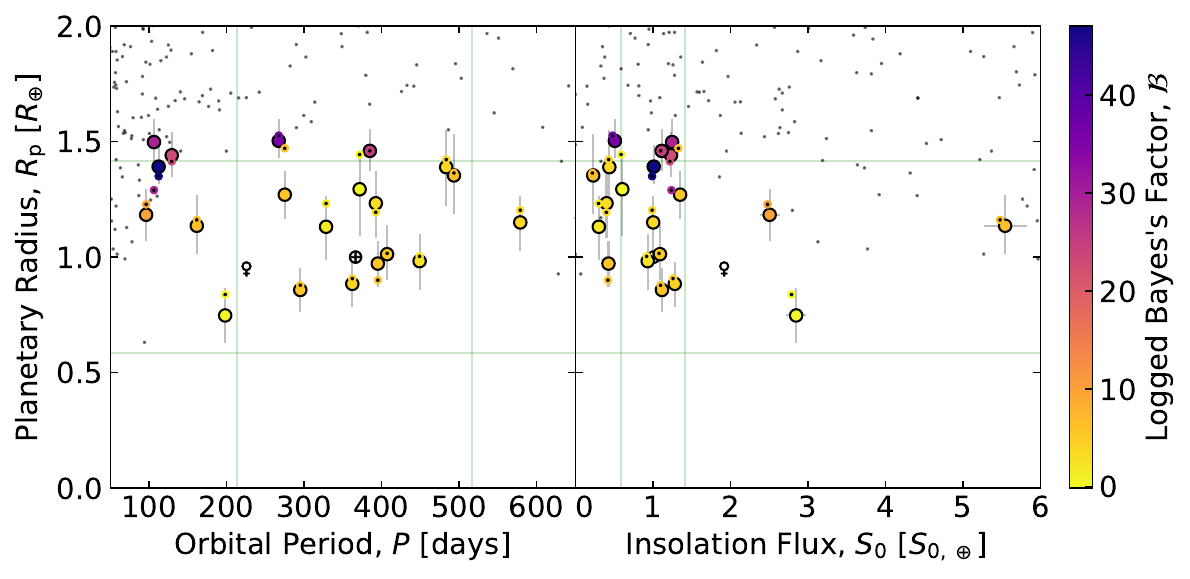}
        \caption[KOI Sample]{Our sample population of \KOI{}s (big colored circles with black outlines) and the remaining \KOI{} background population (small black dots), distributed according to planetary radius, \Rp{}, orbital period, \Period{} (left), and insolation flux, \Szero{} (right). Note that our sample uses the newly fitted/derived results of this work whereas background \KOI{}s draw from the preceding MCMC solutions of \citet{2023_10_Lissauer}. Our \KOI{}s are colored by their logged Bayes's factor, \logB{}---as recovered by the modelling of each individual \KOI{} under \PC{} (transit plus Gaussian process, \TGP{}) and \FA{} (Gaussian process only, \GP{}) hypotheses---such that greater positive values indicate strong planet-candidacy and vice-versa for \FA{}s, while those near-zero can be interpreted as possessing inconclusive/weak significance either way (see \autoref{subsec:Model Comparison}). The \logB{} values corresponding to our solutions are also used to outline associated \citet{2023_10_Lissauer} results (small black dots with colored outlines) in order to facilitate a visual comparison of the \textit{physical} parameters recovered for each \KOI{} analyzed by both studies. Green lines outline the range within which a \KOI{} may be deemed sufficiently ``Earth--Sun-like''; these are defined according to nominal Earth values for \Rp{}, and either \Period{} or \Szero{} as $x \in \left[ 1 \pm \left( \sqrt{2} - 1 \right) \right] x_{\Earth}$. Note that all \KOI{}s were also uniformly filtered by \Rstar{} with respect to solar values according to these same bounds. We are complete in both boxes drawn by these lines. For ease of reference, Earth $\left( \Earth \right)$ and Venus $\left( \Venus \right)$ are also plotted.} 
        \label{fig:KOI Sample}
    \end{figure*}
    
    We describe our proposed methodology herein and apply it to select \ik targets, including potential Earth--Sun and Venus--Sun analogs (see \autoref{fig:KOI Sample}). In \autoref{sec:Methodology} we lay the statistical foundation for Bayesian model comparison, \GP{}s, and nested sampling (\autoref{subsec:Model Comparison}) before proceeding with the construction of our \TGP{} and \GP{} models (\autoref{subsec:Summary of Models}), an overview of the software architecture (\autoref{subsec:Software Architecture}), and ending with a summary of how we obtain derived parameters from fitted solutions (\autoref{subsec:Derived Parameters}). We identify a sample population of small long-period low-\SNR{} \KOI{}s---including \ikt's most Earth--Sun-like exoplanet systems: Kepler-62f \citep[\KOI{}-701.04;][]{2013_05_Borucki}, Kepler-442b \citep[\KOI{}-4742.01;][]{2015_02_Torres}, and Kepler-452b \citep[\KOI{}-7016.01;][]{2015_08_Jenkins}---whose preceding Markov Chain Monte-Carlo \citep[MCMC;][]{1953_06_Metropolis,1970_04_Hastings} solutions indicate potential for Earth--Sun and/or Venus--Sun analog candidacy in \autoref{sec:Candidates Modelled Herein}. The subsections of \autoref{sec:Numerical Results} present and analyze \TGP{} and \GP{} \UltraNest{} solutions relative to each other in the context of Bayesian evidences and potential biases. In \autoref{subsec:Strong and Weak Cases}, we interpret \UltraNest{} solutions of Kepler-62f and \KOI{}-5227.01 to establish expected behavior from strong/weak \PC{}s. The widths of phased photometric data windows and priors have the potential to influence the recovered logged Bayes's factor between models; the effects of this are explored throughout \autoref{subsec:Varying Free Parameters}. This section is closed off by a comparison of the Bayesian evidence against the standard metrics of \MES{} and \SNR{} in \autoref{subsec:Comparisons to Multiple Event Statistic and Signal-to-Noise Ratio}. We conclude with a summary of this paper's leading results in \autoref{sec:Conclusions} and outline future work in \autoref{sec:Next Steps}. A list of terminology and acronyms can be found in \autoref{sec:Terminology and Acronyms}. 

\section[Methodology]{Methodology} \label{sec:Methodology}
    
    In this section, we introduce the reader to fundamental methodology upon which we base our analysis, beginning with a summary of Bayesian statistics and evidence-based model comparison in \autoref{subsec:Model Comparison}. Our combined treatment of white and correlated noise by use of a Gaussian distribution and Mat{\'e}rn 3/2 kernel \GP{} is established next. Following this, \autoref{subsec:Summary of Models} provides a breakdown of each model (\TGP{} and \GP{}) in terms of their parameters. A step-by-step outline of our \UltraNest{} software architecture and model fitting process for any given \KOI{} can be read in \autoref{subsec:Software Architecture}. Derived parameter calculations are detailed in \autoref{subsec:Derived Parameters}.

\subsection[Model Comparison]{Model Comparison} \label{subsec:Model Comparison}

    Bayes's theorem \citep{1763_01_Bayes,1774_Laplace}---which forms the basis of Bayesian statistics and probability theory---describes the process by which our knowledge of an event (posterior) is probabilistically updated according to existing information (prior) and new observations (likelihood). In other words, it allows us to adjust our understanding of the world in order to make better informed decisions/predictions. From a statistical perspective, we can model observed data, \data{}, via the inference of model parameters, \parameters{}, using Bayes's theorem:
    \begin{equation} \label{eq:Bayes's Theorem}
        \posteriors{} = \frac{\likelihood{} \priors{}}{\evidence{}} \ ,
    \end{equation}
    where the posteriors, \posteriors{}, are represented in terms of the likelihood, \likelihood{}, priors, \priors{}, and Bayesian evidence (marginal likelihood integral), \evidence{}. Here, \evidence{} gives the probability associated with observing this realization of \data{} and is defined as:
    \begin{equation} \label{eq:Evidence Integral}
        \evidence{} \equiv \int \likelihood{} \priors{} d\parameters{} \ .
    \end{equation} 
    The \evidence{} encodes both \likelihood{} and \priors{} information, so it is often employed as a metric of model suitability. Should one know the most suitable model for a given problem, the computationally expensive \evidence{} can be readily discarded in favor of obtaining only \posteriors{} of modelled \parameters{} (e.g., likelihood-driven techniques such as MCMC). However, it is uncommon in real-world problems to possess the most suitable model with which \data{} is described in totality. As such, \evidence{}, and by extension the Bayes's factor of any two models, $A$ and $B$,
    \begin{equation} \label{eq:Bayes's Factor}
        \mathcal{B}_{A , B} \equiv \frac{\evidence{}_{, A}}{\evidence{}_{, B}} \ ,
    \end{equation}
    play a crucial role in determining the most suitable model for \data{}. This statistically robust process of model selection is known as Bayesian model comparison \citep{1939_Jeffreys,1995_06_Kass,2003_Mackay,2020_07_Dunstan,2022_01_Dunstan}.
    Our study applies the logged form of the Bayes's factor:
    \begin{equation} \label{eq:Logged Bayes's Factor}
        \log \left( \mathcal{B}_{A , B} \right) = \log \left( \evidence{}_{, A} \right) - \log \left( \evidence{}_{, B} \right) \ ;
    \end{equation}
    this quantity is defined such that positive values favor model $A$ and negative values model $B$. We replace $A$ by the transit plus Gaussian process (\TGP{}; \PC{} hypothesis) model and $B$ by the Gaussian process (\GP{}; \FA{} hypothesis) model to obtain $\log \left( \mathcal{B}_{\rm TGP , GP} \right)$---whose notation we condense to \logB{}:
    \begin{equation} \label{eq:logB}
        \logB{} \equiv \log \left( \mathcal{B}_{\rm TGP , GP} \right) \ .
    \end{equation}
    Accordingly, increasingly positive values of \logB{} promote the existence of the transiting \PC{} while their negative counterparts suggest a \FA{} signal originating purely from noise. Values of \logB{} near zero indicate no statistically significant improvement given by the addition of transit parameters to the fit with respect to the null (noise) hypothesis; that is not to say that these are definitively \PC{}s or \FA{}s or that either fit is necessarily less robust, but that no statistically significant difference exists between hypotheses.

    Given an informed choice of kernel \citep[covariance function; see][]{2006_Rasmussen}, a \GP{} may target specific behavior or systematics within a given data set---this is particularly useful when attempting to fit correlated noise present in photometric time series observations. The Squared-Exponential \citep[Radial Basis Function;][]{1966_Cheney,1975_Davis,1981_Powell} and Mat{\'e}rn \citep{1960_Matern} kernel families have become popular for the treatment of systematics in astronomy \citep[][etc.]{2012_01_Gibson,2012_12_Roberts,2015_02_Barclay,2015_03_Aigrain,2015_07_Evans,2015_10_Czekala,2016_07_Aigrain,2017_04_Littlefair,2017_12_Foreman_Mackey,2018_02_Angus,2019_03_Livingston,2023_06_Brahm,2023_08_Aigrain}; while both are generally well-suited to smooth signal applications (e.g., stellar variability), the latter is also capable of handling rougher interference \citep[e.g., sudden pixel sensitivity dropout as illustrated by Figure~19 of][]{2018_04_Thompson}. Given the known characteristics of stellar and instrumental \FA{} sources which contaminate \ik photometry \citep{2011_11_Gilliland,2015_10_Gilliland,2016_04_Van_Cleve,2016_12_Van_Cleve,2018_04_Thompson}, we adopt the Mat{\'e}rn 3/2 kernel:
    \begin{equation} \label{eq:Matern 3/2}
        k \equiv \sigmac{}^{2} \left( 1 + \frac{\sqrt{3} \left\vert t - t^{\prime} \right\vert}{\lc{}} \right) \exp \left( - \frac{\sqrt{3} \left\vert t - t^{\prime} \right\vert}{\lc{}} \right) \ .
    \end{equation} 
    Here, \sigmac{} and \lc{} describe the amplitude and length scales of the correlated noise with which every pair of data points, $t$ and $t^{\prime}$, is conditioned. White noise is incorporated as a scaling factor to the error bars belonging to each photometric observation and is obtained by fitting the standard deviation, \sigmaw{}, of a zero-mean Gaussian.
    
    Nested sampling is a popular class of algorithm which approximates \autoref{eq:Evidence Integral} and provides posterior inference(s) as byproducts given \data{}, \likelihood{}, and \priors{}. Our current infrastructure makes use of \UltraNest{}, which requires user-defined \likelihood{},
    and prior transforms or quantile functions mapping between physical parameter and unit hypercube sampling spaces. Uniform priors are used for all \TGP{}/\GP{} parameters excluding limb-darkening parameters, \qone{} and \qtwo{}, which instead use Gaussian priors \citep{2013_11_Kipping}.

\subsection[Summary of Models]{Summary of Models} \label{subsec:Summary of Models}

    \begin{table}[htb!]
        \centering
        \caption[Model Parameter Descriptions]{Model Parameter Descriptions} 
        \label{tab:Model Parameter Descriptions}
        \begin{tabular}{p{0.98\hsize}}
            \textbf{Transit Model Parameters} \\
            \hline
            $\bullet$ \rhostar{}: Mean stellar density $\left[ \gcmthree{} \right]$. \\
            $\bullet$ \qone{}: \citet{2013_11_Kipping} limb-darkening [unitless]. \\
            $\bullet$ \qtwo{}: \citet{2013_11_Kipping} limb-darkening [unitless]. \\
            $\bullet$ \Tzero{}: Transit time series epicenter $\left[ \days{} \right]$. \\
            $\bullet$ \Period{}: Orbital period of the exoplanet $\left[ \days{} \right]$. \\
            $\bullet$ \bimpact{}: Impact parameter [unitless]. \\
            $\bullet$ \RpRstar{}: Ratio of planetary and stellar radii [unitless]. \\
            $\bullet$ \Fzero{}: Relative photometric zero-point offset [unitless]. \\
            \textbf{Noise Model Parameters} \\
            \hline
            $\bullet$ \sigmaw{}: Multiplicative factor applied to the photometric errors reported by DR25 (white noise) [unitless]. \\
            $\bullet$ \sigmac{}: Amplitude scale of \autoref{eq:Matern 3/2} [unitless]. \\
            $\bullet$ \lc{}: Length scale of \autoref{eq:Matern 3/2} [unitless].
        \end{tabular}
    \end{table}

    Photometric exoplanet transits were modeled using \transitfitfive{} \citep{2016_08_Rowe}. The lightcurve model uses the analytic limb-darkening transit from \citet{2002_12_Mandel} and assumes non-interacting Keplerian orbits. The model is parameterized with \rhostar{}, \qone{}, \qtwo{}, \Tzero{}, \Period{}, \bimpact{}, \RpRstar{}, \Fzero{}, \ecosw{}, \esinw{}, and a photometric dilution factor (see \autoref{tab:Model Parameter Descriptions}). The model can additionally include the effects of geometric albedo, ellipsoidal variations, and secondary eclipses. The calculation of Keplerian orbits derives the scaled semi-major axis, \aRstar{}, based on \rhostar{}; this calculation assumes that the planetary mass, \Mp{}, is much less than the mass of the host star, \Mstar{}. For all presented models in this paper, we assume: (1) circular orbits (i.e., zero eccentricity, \eccentricity{}) such that $\ecosw{} = \esinw{} = 0$, (2) no dilution\footnote{DR25 lightcurves already include a crowding correction for other stars that contribute to the photometric aperture.} (unresolved binaries), (3) no star-planet interactions, and (4) that the planet is completely dark (no reflection or emission). Limb-darkening parameters are from the tables of \citet{2011_05_Claret} for the \ik bandpass. \rev{The shape information of low-\SNR{} putative transit signatures within our regime of interest: (1) leaves \eccentricity{} and \wperi{} very weakly constrained and (2) makes uninformative limb-darkening priors superfluous.} These assumptions and inputs to the modelling approach are similar to transit model results presented in DR25. 
        
\subsection[Software Architecture]{Software Architecture} \label{subsec:Software Architecture}

    The general step-by-step outline for the fitting of an individual \KOI{} is detailed in this section. 

    \begin{enumerate}
        \item Since the preceding MCMC architecture of \citet{2023_10_Lissauer} modelled the transit events of prewhitened data rather than simultaneously fitting correlated noise and transit events as performed in this study, we treat their transit solutions as initial guesses to define focused prior widths for our transit model in \UltraNest{}; this mitigates computationally wasteful exploration of uninformative/unlikely parameter space. 
        \item For noise model hyperparameters, we define physically-motivated prior widths accordingly: 
            \begin{enumerate}
                \item While \ik photometry typically falls within 10~--~20\% of the predicted white noise budget \citep{2011_11_Gilliland,2015_10_Gilliland,2016_04_Van_Cleve,2016_12_Van_Cleve,2018_04_Thompson}, we err on the side of caution with a wide uninformative prior on the scaled photometric error of $\sigmaw{} \in \left[ 0.2 , 3.8 \right]$. 
                \item The amplitude scale of the correlated noise has prior width set as $\sigmac{} \in \left[ 0 , \left( F_{max} - F_{min} \right) / 2 \right]$; this should not exceed the maximum flux semi-amplitude, $\left( F_{max} - F_{min} \right) / 2$, observed in a given \KOI{}'s data set. 
                \item The length scale of the correlated noise has prior width set as $\lc{} \in \left[ 2 \Deltat{} , 2 \sigmad{} \tT{} \right]$; this should not fall below the measurement cadence, \Deltat{}, or exceed the phased photometric data window's timescale, $2 \sigmad{} \tT{}$. 
            \end{enumerate}
        \item Define likelihood and prior cube transformation functions for \UltraNest{}.
        \item Set free parameters: \sigmad{} and \sigmap{}.
        \item Initialize and precompute all relevant values (i.e., \GP{} kernel). 
        \item Conduct photometric time series data preprocessing/reduction, including the following:
        \begin{enumerate}
            \item Removal of other threshold-crossing events (\TCE{}s)/\KOI{}s associated with the same host star.
            \item \sigmad{} phased photometric data window width specification.
            \item Linear regression removal of ramps/slopes.
            \item Median-based zero-point correction.
            \item Removal of data $\pm 5 \sigma$ beyond the MCMC data-residual median value to deal with uncorrected cosmic rays, flares, or uncorrected instrumental effects following \citet{2018_04_Thompson}. 
        \end{enumerate}
        \item Run \UltraNest{}'s \ReactiveNestedSampler{} with \RegionSliceSampler{} enabled once per model (\TGP{} and \GP{}). 
    \end{enumerate}

    To solve for the Mat{\'e}rn 3/2 \GP{}'s hyperparameters prior to each iteration's likelihood evaluation, matrix inversion must be performed. Since the transit events are effectively isolated in time, their correlated noise components can be approximated to share negligible covariance; we represent this by means of a block-diagonal approximation to the kernel, drastically decreasing the computational burden of matrix inversions. Naturally, benefits in performance scale with the number of transit events in a given \KOI{}'s data set. For the task of inversion, we use Cholesky decomposition---a method roughly twice as efficient as lower-upper (LU) decomposition \citep{1924_04_Cholesky,1942_08_Banachiewicz,1986_Press,1995_04_Schwarzenberg_Czerny}. Nonetheless, each iteration is still expensive. 

\subsection[Derived Parameters]{Derived Parameters} \label{subsec:Derived Parameters}

    Formulas for derived parameters can be found within this section. We compute transit duration according to Equation 16 of \citet{2003_03_Seager}, rewritten here using Kepler's Third Law \citep{1619_Kepler} as: 
    \begin{equation} \label{eq:Transit Duration rho}
        \tT{} = \left( \frac{3 \Period{}}{\pi^{2} G \rhostar{}} \right)^{\frac{1}{3}} \sqrt{\left( 1 + \frac{\Rp{}}{\Rstar{}} \right)^{2} - \bimpact{}^{2}}.
    \end{equation}
    
    The insolation flux, $\Szero{} \equiv L / 4 \pi a^{2}$,
    can be combined with Kepler's Third Law to yield:
    \begin{equation} \label{eq:Insolation Flux}
        \frac{\Szero{}}{\SzeroEarth{}} = \frac{\Lstar{}}{\LSun{}} \left[ \frac{\Mstar{}}{\MSun{}} \left( \frac{\Period{}}{\PeriodEarth{}} \right)^{2} \right]^{- \frac{2}{3}} \ .
    \end{equation}
    Isochrone-derived stellar luminosity and mass from \citet{2020_06_Berger} are used alongside our fitted orbital period to compute insolation flux via \autoref{eq:Insolation Flux}.
    
    \rev{Although we do not model eccentricity directly \citep[e.g.,][]{2019_02_Van_Eylen}}, a minimum eccentricity may be estimated following \citet{2014_05_Kipping} and \citet{2015_02_Torres} via the comparison of stellar density recovered by our model against independent estimates:
    \begin{equation} \label{eq:Minimum Eccentricity}
        \eccentricitymin{} \equiv \frac{\left\vert 1 - \left( \rhostar{}_{\rm , model} / \rhostar{}_{\rm , indep} \right)^{2 / 3} \right\vert}{1 + \left( \rhostar{}_{\rm , model} / \rhostar{}_{\rm , indep} \right)^{2 / 3}} \ .
    \end{equation}

    To obtain distributions of independent parameters, we use the reported value and lower/upper uncertainties to sample from a two-piece normal distribution \citep{2014_05_Wallis}. For example, the \Rp{} distribution is derived via a convolution between the fitted \RpRstar{} posterior distribution and a two-piece Gaussian distribution of \Rstar{} constructed from \citet{2020_06_Berger}.
    
\section[Candidates Modelled Herein]{Candidates Modelled Herein} \label{sec:Candidates Modelled Herein}

    Targets were selected from a comprehensive catalog of \ik candidates with revised light curve analyses \citep{2023_10_Lissauer}. Included within our sample are 12 \PC{}s whose planetary and host stellar radii, \Rp{} and \Rstar{}, plus either orbital period, \Period{}, or insolation flux, \Szero{}, nominally (neglecting uncertainties) lie within $\pm \left( \sqrt{2} - 1 \right)$ of Earth and solar values $\left(\text{e.g., }\Rp{} \in \left[ 1 \pm \left( \sqrt{2} - 1 \right) \right] \REarth{} \right)$. There are just 13 \KOI{}s identified by these criteria: 2719.02, 4742.01, 4878.01, 5554.01, 5755.01, 5971.01, 6971.01, 7179.01, 7470.01, 7591.01, 7923.01, 8107.01, and 8174.01; we have not \textit{yet} analyzed \KOI{}-5755.01 because it lacks a converged preceding MCMC solution. 
    
    The nature of this regime places candidates at significant risk of being \FA{}s; $\MES{} \lesssim 8$ and $\SNR{} \lesssim 10$ are predominantly exhibited within our sample. Furthermore, \KOI{}s-5044.01, 5971.01, 7621.01, and 7923.01 yield suspect derived transit durations, \tT{}, which diverge significantly from those expected for equatorial transits of planets on circular orbits, \tTc{}; regardless, similar performance here between \TGP{} and \GP{} models (or \PC{} and \FA{} hypotheses) has resulted in none of these \KOI{}s possessing high \logB{}. 
    
    In addition to those candidates listed above, we also included other small long-period \PC{}s; among these are three of \ikt's validated exoplanets whose characteristics most closely approach the Earth--Sun analog regime---excellent targets against which we may baseline our framework---: Kepler-62f (\KOI{}-701.04), Kepler-442b (\KOI{}-4742.01), and Kepler-452b (\KOI{}-7016.01). The comprehensive target list is given in \autoref{tab:Key Parameters Summary} alongside fitted and derived parameters, as well as \citet{2020_06_Berger} stellar property inputs for the free parameter choice of $\sigmad{} = 8$ and $\sigmap{} = 5$. 

    \begin{table*}[htb!]
        \centering
        \caption[Key Parameters Summary]{Key Parameters Summary}
        \label{tab:Key Parameters Summary}
        \begin{tabular}{ccccccccccccc}
            $\KOI{}$ & $\boldsymbol{\Period{}}$ & $\boldsymbol{\tT{}}$ & $\boldsymbol{\tTc{}}$ & $\boldsymbol{\Szero{}}$ & $\boldsymbol{\Rp{}}$ & \Rpprime{} & \Rstar{} & \Teff{} & \MES{} & \SNRprime{} & $\boldsymbol{\SNR{}}$ & $\boldsymbol{\logB{}}$ \\ 
            ~ & $\boldsymbol{\left[ \days{} \right]}$ & $\boldsymbol{\left[ \hours{} \right]}$ & $\boldsymbol{\left[ \hours{} \right]}$ & $\boldsymbol{\left[ \SzeroEarth{} \right]}$ & $\boldsymbol{\left[ \REarth{} \right]}$ & $\left[ \REarth{} \right]$ & $\left[ \RSun{} \right]$ & $\left[ \Kelvin{} \right]$ & ~ & ~ & ~ & ~ \\ 
            \hline
            701.04 & $\boldsymbol{267.282}$ & $\boldsymbol{7.8_{- 0.3}^{+ 0.3}}$ & $\boldsymbol{7.7_{- 1.7}^{+ 4.1}}$ & $\boldsymbol{0.51}$ & $\boldsymbol{1.50_{- 0.07}^{+ 0.10}}$ & $1.52_{- 0.11}^{+ 0.22}$ & $0.70$ & $4967$ & $14.3$ & $19$ & $\boldsymbol{19_{- 2}^{+ 2}}$ & $\boldsymbol{35.8}$ \\ 
            2719.02 & $\boldsymbol{106.261}$ & $\boldsymbol{6.6_{- 0.3}^{+ 0.3}}$ & $\boldsymbol{7.6_{- 0.2}^{+ 0.2}}$ & $\boldsymbol{1.25}$ & $\boldsymbol{1.50_{- 0.10}^{+ 0.10}}$ & $1.29_{- 0.13}^{+ 0.16}$ & $0.69$ & $4601$ & $10.0$ & $7.7$ & $\boldsymbol{15_{- 2}^{+ 2}}$ & $\boldsymbol{29.4}$ \\ 
            4742.01 & $\boldsymbol{112.303}$ & $\boldsymbol{5.9_{- 0.3}^{+ 0.3}}$ & $\boldsymbol{7.2_{- 0.1}^{+ 0.1}}$ & $\boldsymbol{1.01}$ & $\boldsymbol{1.39_{- 0.08}^{+ 0.09}}$ & $1.35_{- 0.10}^{+ 0.12}$ & $0.62$ & $4602$ & $12.9$ & $13$ & $\boldsymbol{14_{- 1}^{+ 1}}$ & $\boldsymbol{47.1}$ \\ 
            4878.01 & $\boldsymbol{449.016}$ & $\boldsymbol{13.4_{- 1.4}^{+ 2.1}}$ & $\boldsymbol{16.6_{- 0.6}^{+ 0.6}}$ & $\boldsymbol{0.93}$ & $\boldsymbol{0.98_{- 0.12}^{+ 0.12}}$ & $1.01_{- 0.10}^{+ 0.11}$ & $1.05$ & $5906$ & $7.5^{*}$ & $8.3$ & $\boldsymbol{8.0_{- 1.9}^{+ 1.8}}$ & $\boldsymbol{2.1}$ \\ 
            5044.01 & $\boldsymbol{161.533}$ & $\boldsymbol{2.1_{- 0.5}^{+ 0.4}}$ & $\boldsymbol{12.7_{- 0.4}^{+ 0.5}}$ & $\boldsymbol{5.55}$ & $\boldsymbol{1.14_{- 0.12}^{+ 0.13}}$ & $1.16_{- 0.15}^{+ 0.16}$ & $1.11$ & $6344$ & $8.4^{*}$ & $6.9$ & $\boldsymbol{5.9_{- 1.1}^{+ 1.1}}$ & $\boldsymbol{6.4}$ \\ 
            5227.01 & $\boldsymbol{371.653}$ & $\boldsymbol{11.2_{- 1.8}^{+ 3.1}}$ & $\boldsymbol{12.9_{- 0.3}^{+ 0.4}}$ & $\boldsymbol{0.61}$ & $\boldsymbol{1.29_{- 0.20}^{+ 0.17}}$ & $1.45_{- 0.11}^{+ 0.15}$ & $0.83$ & $5486$ & $8.4^{*}$ & $11$ & $\boldsymbol{8.2_{- 2.6}^{+ 2.5}}$ & $\boldsymbol{- 0.1}$ \\ 
            5554.01 & $\boldsymbol{362.181}$ & $\boldsymbol{18.7_{- 2.6}^{+ 3.5}}$ & $\boldsymbol{15.5_{- 0.5}^{+ 0.6}}$ & $\boldsymbol{1.28}$ & $\boldsymbol{0.88_{- 0.10}^{+ 0.09}}$ & $0.91_{- 0.07}^{+ 0.07}$ & $1.04$ & $5945$ & $7.3^{*}$ & $12$ & $\boldsymbol{13_{- 3}^{+ 3}}$ & $\boldsymbol{4.1}$ \\ 
            5704.01 & $\boldsymbol{96.167}$ & $\boldsymbol{2.7_{- 0.3}^{+ 0.3}}$ & $\boldsymbol{7.9_{- 0.2}^{+ 0.2}}$ & $\boldsymbol{2.51}$ & $\boldsymbol{1.18_{- 0.11}^{+ 0.11}}$ & $1.23_{- 0.17}^{+ 0.15}$ & $0.75$ & $5075$ & $7.6^{*}$ & $7.7$ & $\boldsymbol{6.9_{- 1.2}^{+ 1.1}}$ & $\boldsymbol{9.8}$ \\ 
            5971.01 & $\boldsymbol{493.328}$ & $\boldsymbol{3.1_{- 0.8}^{+ 0.9}}$ & $\boldsymbol{12.0_{- 0.4}^{+ 0.5}}$ & $\boldsymbol{0.23}$ & $\boldsymbol{1.35_{- 0.17}^{+ 0.18}}$ & $1.36_{- 0.15}^{+ 0.22}$ & $0.76$ & $4847$ & $7.6^{*}$ & $7.8$ & $\boldsymbol{5.9_{- 1.1}^{+ 1.1}}$ & $\boldsymbol{6.3}$ \\ 
            6971.01 & $\boldsymbol{129.222}$ & $\boldsymbol{6.9_{- 0.4}^{+ 0.4}}$ & $\boldsymbol{8.0_{- 0.2}^{+ 0.2}}$ & $\boldsymbol{1.23}$ & $\boldsymbol{1.44_{- 0.09}^{+ 0.10}}$ & $1.42_{- 0.11}^{+ 0.13}$ & $0.68$ & $4921$ & $8.1$ & $12$ & $\boldsymbol{14_{- 2}^{+ 2}}$ & $\boldsymbol{22.6}$ \\ 
            7016.01 & $\boldsymbol{384.847}$ & $\boldsymbol{10.2_{- 0.5}^{+ 0.5}}$ & $\boldsymbol{15.7_{- 0.4}^{+ 0.5}}$ & $\boldsymbol{1.11}$ & $\boldsymbol{1.46_{- 0.09}^{+ 0.09}}$ & $1.46_{- 0.09}^{+ 0.13}$ & $1.07$ & $5900$ & $7.6$ & $12$ & $\boldsymbol{12_{- 1}^{+ 1}}$ & $\boldsymbol{24.9}$ \\ 
            7179.01 & $\boldsymbol{407.093}$ & $\boldsymbol{13.9_{- 5.1}^{+ 2.6}}$ & $\boldsymbol{16.1_{- 0.5}^{+ 0.6}}$ & $\boldsymbol{1.09}$ & $\boldsymbol{1.01_{- 0.11}^{+ 0.13}}$ & $1.02_{- 0.11}^{+ 0.11}$ & $1.06$ & $5946$ & $7.8$ & $8.4$ & $\boldsymbol{7.4_{- 1.8}^{+ 1.7}}$ & $\boldsymbol{5.2}$ \\ 
            7330.01 & $\boldsymbol{198.141}$ & $\boldsymbol{12.2_{- 3.1}^{+ 3.6}}$ & $\boldsymbol{14.7_{- 0.5}^{+ 0.5}}$ & $\boldsymbol{2.85}$ & $\boldsymbol{0.75_{- 0.12}^{+ 0.12}}$ & $0.84_{- 0.09}^{+ 0.08}$ & $1.19$ & $5490$ & $8.0$ & $7.2$ & $\boldsymbol{6.0_{- 1.7}^{+ 1.9}}$ & $\boldsymbol{0.2}$ \\ 
            7470.01 & $\boldsymbol{392.553}$ & $\boldsymbol{8.2_{- 1.4}^{+ 1.4}}$ & $\boldsymbol{13.3_{- 0.4}^{+ 0.4}}$ & $\boldsymbol{0.40}$ & $\boldsymbol{1.23_{- 0.15}^{+ 0.14}}$ & $1.20_{- 0.15}^{+ 0.16}$ & $0.82$ & $5016$ & $7.2$ & $7.2$ & $\boldsymbol{6.9_{- 1.5}^{+ 1.6}}$ & $\boldsymbol{3.0}$ \\ 
            7591.01 & $\boldsymbol{328.339}$ & $\boldsymbol{15.7_{- 4.3}^{+ 6.1}}$ & $\boldsymbol{10.6_{- 0.2}^{+ 0.2}}$ & $\boldsymbol{0.30}$ & $\boldsymbol{1.13_{- 0.14}^{+ 0.13}}$ & $1.24_{- 0.16}^{+ 0.12}$ & $0.65$ & $4786$ & $7.4$ & $8.5$ & $\boldsymbol{7.5_{- 1.9}^{+ 2.1}}$ & $\boldsymbol{1.9}$ \\ 
            7621.01 & $\boldsymbol{275.075}$ & $\boldsymbol{2.9_{- 0.2}^{+ 0.2}}$ & $\boldsymbol{13.4_{- 0.4}^{+ 0.5}}$ & $\boldsymbol{1.35}$ & $\boldsymbol{1.27_{- 0.11}^{+ 0.10}}$ & $1.47_{- 0.22}^{+ 0.18}$ & $0.97$ & $5662$ & $8.0$ & $7.5$ & $\boldsymbol{18_{- 3}^{+ 3}}$ & $\boldsymbol{5.7}$ \\ 
            7716.01 & $\boldsymbol{483.346}$ & $\boldsymbol{15.8_{- 1.9}^{+ 2.6}}$ & $\boldsymbol{14.5_{- 0.6}^{+ 0.8}}$ & $\boldsymbol{0.44}$ & $\boldsymbol{1.39_{- 0.17}^{+ 0.16}}$ & $1.42_{- 0.15}^{+ 0.19}$ & $0.85$ & $5451$ & $7.1$ & $8.3$ & $\boldsymbol{7.7_{- 1.8}^{+ 1.7}}$ & $\boldsymbol{3.5}$ \\ 
            7923.01 & $\boldsymbol{395.122}$ & $\boldsymbol{21.4_{- 0.8}^{+ 0.9}}$ & $\boldsymbol{13.5_{- 0.2}^{+ 0.2}}$ & $\boldsymbol{0.43}$ & $\boldsymbol{0.97_{- 0.10}^{+ 0.10}}$ & $0.90_{- 0.07}^{+ 0.12}$ & $0.82$ & $5064$ & $10.0$ & $14$ & $\boldsymbol{19_{- 4}^{+ 3}}$ & $\boldsymbol{5.6}$ \\ 
            8107.01 & $\boldsymbol{578.916}$ & $\boldsymbol{18.8_{- 0.9}^{+ 1.5}}$ & $\boldsymbol{22.8_{- 0.8}^{+ 0.9}}$ & $\boldsymbol{1.00}$ & $\boldsymbol{1.15_{- 0.12}^{+ 0.12}}$ & $1.20_{- 0.07}^{+ 0.09}$ & $1.35$ & $5832$ & $7.6$ & $15$ & $\boldsymbol{14_{- 3}^{+ 3}}$ & $\boldsymbol{3.5}$ \\ 
            8174.01 & $\boldsymbol{295.061}$ & $\boldsymbol{16.5_{- 1.5}^{+ 1.3}}$ & $\boldsymbol{14.9_{- 0.4}^{+ 0.4}}$ & $\boldsymbol{1.12}$ & $\boldsymbol{0.86_{- 0.09}^{+ 0.10}}$ & $0.88_{- 0.07}^{+ 0.09}$ & $1.04$ & $5284$ & $7.4$ & $10$ & $\boldsymbol{11_{- 2}^{+ 2}}$ & $\boldsymbol{5.7}$ \\ 
        \end{tabular}
        \caption*{Key parameters summary for our \KOI{} sample. From left to right are \KOI{} numbers (\KOI{}-701.04, \KOI{}-4742.01, and \KOI{}-7016.01 correspond to Kepler-62f, Kepler-442b, and Kepler-452b, respectively; none of the other \KOI{}s are validated \ik planets), orbital period, \Period{}, transit duration, \tT{}, central transit duration, \tTc{}, insolation flux, \Szero{}, planetary and stellar radii, \Rp{} and \Rstar{}, stellar effective temperature, \Teff{}, Multiple Event Statistic, \MES{}, signal-to-noise, \SNR{}, and logged Bayes's factor, \logB{}. Of these, the \Period{} and \logB{} are fitted (inputs), \tT{}, \tTc{}, \Szero{}, \Rp{}, and \SNR{} are derived (outputs), \Rstar{} and \Teff{} are given by \citet{2020_06_Berger}, \Rpprime{} and \SNRprime{} are given by \citet{2023_10_Lissauer}, and \MES{} is given by DR25 where available; those \KOI{}s not found by DR25 were also missed in Data Release 24, so their \MES{} are instead taken from Q1~--~16 \citep{2015_04_Mullally} and identified using ``$^{*}$''. All fitted/derived parameters are in bold. These values are reported for the free parameter choice of $\sigmad{} = 8$ and $\sigmap{} = 5$.}
    \end{table*}

\newpage
    
\section[Numerical Results]{Numerical Results} \label{sec:Numerical Results}

    There are three leading results to be discussed in this section, beginning with an overview of recovered \TGP{} and \GP{} solutions for cases of strong and weak \PC{} evidence in \autoref{subsec:Strong and Weak Cases}. Here, the former is demonstrated by the baseline target, Kepler-62f, and the latter by a member of our sample \KOI{} population, \KOI{}-5227.01. This is followed by \autoref{subsec:Varying Free Parameters}, which investigates the influence that varying the free model parameters, \sigmad{} and \sigmap{}, has on the \logB{}; these control phased photometric data window and prior widths, respectively. We conclude with a heuristic evaluation of the fitted \logB{} against reported DR25 (or Q1~--~16) \MES{} and derived \SNR{} scores for our target population in \autoref{subsec:Comparisons to Multiple Event Statistic and Signal-to-Noise Ratio}. A summary of results for the complete \KOI{} sample can be found in \autoref{fig:KOI Sample} and \autoref{tab:Key Parameters Summary}.

\subsection[Strong and Weak Cases]{Strong and Weak Cases} \label{subsec:Strong and Weak Cases}

\begin{figure*}[htb!]
        \centering
        \includegraphics[width = \hsize]{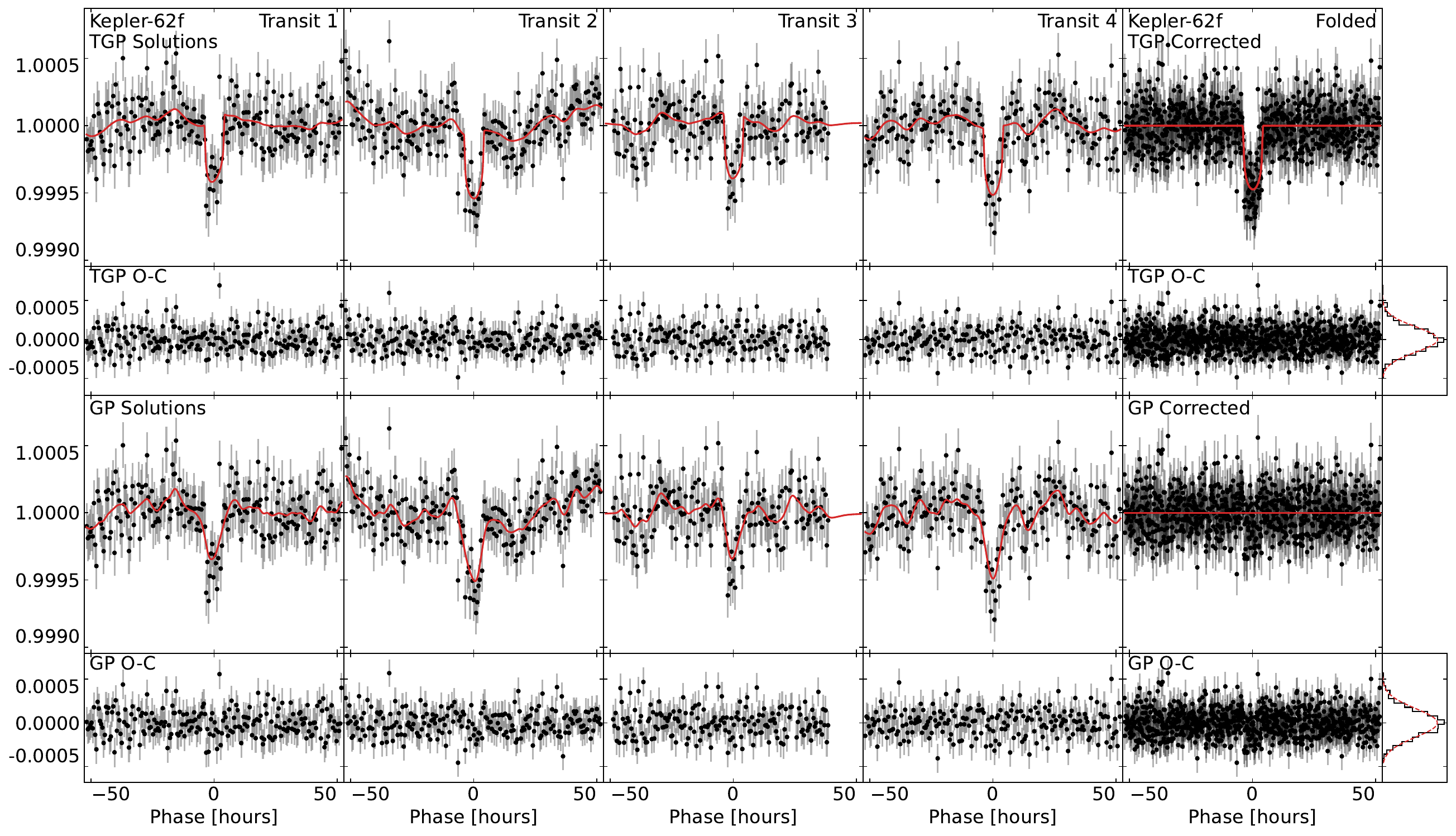}
        \caption[Kepler-62f Phase Plots]{\ikt's photometric PDC data (black) for all observed transit events of Kepler-62f with $\sigmad{} = 8$ and $\sigmap{} = 5$, overlaid by \MAP{} (red) \TGP{} (\PC{} hypothesis; top) and \GP{} (\FA{} hypothesis; bottom) model solutions alongside corresponding O-C (data-\MAP{}) residuals. \GP{}-corrected phase-folded results are shown in the rightmost column and are accompanied by O-C residuals. O-C residual histograms are overlaid by Gaussian distributions with zero mean and \TGP{} or \GP{} \MAP{}-scaled median photometric error standard deviations (dashed red) in order to help identify signs of overfitting (i.e., non-Gaussian O-C residuals).}
        \label{fig:Kepler-62f Phase Plots}
    \end{figure*}

    For the demonstration of strong and weak \PC{} evidence, we compare the recovered \UltraNest{} \TGP{} and \GP{} solutions given free parameter choices of $\sigmad{} = 8$ and $\sigmap{} = 5$ for Kepler-62f and \KOI{}-5227.01. \autoref{fig:Kepler-62f Phase Plots} shows the photometric data---preprocessed according to \autoref{subsec:Software Architecture}---overlaid by \TGP{} and \GP{} solutions in unfolded original and folded \GP{}-corrected states. Although we know Kepler-62f to be a bona~fide exoplanet, both the phase (\autoref{fig:Kepler-62f Phase Plots}) and corner (\autoref{fig:Kepler-62f Corner Plots}) plots demonstrate competitive performance between \PC{} and \FA{} hypotheses; similar model performance is shown for \KOI{}-5227.01 (see \autoref{fig:KOI-5227.01 Phase Plots} and \autoref{fig:KOI-5227.01 Corner Plots}). While it is nontrivial to individually discriminate or relatively rank Kepler-62f and \KOI{}-5227.01 by eye, our statistical analysis places them among the strongest and weakest \PC{}s of the $\sigmad{} = 8$ and $\sigmap{} = 5$ subset, with recovered \logB{} values of $35.8_{- 1.1}^{+ 1.2}$ and $- 0.1_{- 1.2}^{+ 1.1}$, respectively. This translates to strong favor for the former's \PC{} status whereas the latter can be said to fall within the range of values which we deem indistinguishable in that sufficiently strong evidence supporting either hypothesis is lacking. Our example also serves to highlight the importance of both joint transit-noise modelling and robust Bayesian model comparison techniques---especially when working within lower signal-to-noise regimes such as these---, without which we would be more susceptible to target misclassification. 
    
    Generally, we have seen (\autoref{fig:Data Window Variance Results} and \autoref{fig:Prior Width Variance Results}) two sub-populations reflective of the above comparison emerge from our analysis (strong \PC{}s and inconclusive/weak \PC{}s and/or \FA{}s). Included in the strong \PC{} group are reassuringly three known exoplanets, Kepler-62f, Kepler-442b, and Kepler-452b, in addition to two more promising but less potentially Earth--Sun-like \PC{}s, \KOI{}-2719.02 and \KOI{}-6971.01. The remainder of \autoref{tab:Key Parameters Summary}'s candidates fall within the currently indistinguishable range of $\sim \pm 10$ \logB{}; further observations and/or deeper probabilistic analyses are likely required before more definitive conclusions may be made. 
    
    It should be noted that we do not yet observe candidates whose \logB{} strongly favors the \FA{} hypothesis. We suspect this to be the result of: (1) survivor bias potentially introduced by our lack of targets with \MES{} or \SNR{} below $\sim 7$ (i.e., we have yet to analyze sufficiently poor photometry) and/or (2) a miscalibrated \logB{} scale (i.e., the magnitude of the \evidence{} penalty incurred by the \TGP{} model's additional parameters and by extension, the \logB{} floor, are currently unknown). To obtain statistically robust conclusions to these hypotheses, future work will implement large-scale injection-recovery testing.

\subsection[Varying Free Parameters]{Varying Free Parameters} \label{subsec:Varying Free Parameters}

    There are two free parameters required to initialize our modelling pipeline, these being \sigmad{} and \sigmap{}. Of concern to us is their influence on the recovered posteriors and \logB{}. We begin with \sigmad{}, which acts as a multiplicative factor to the width of the phased photometric data window, defined as $2 \sigmad{} \tT{}$, where $\tT{}$ is the transit duration as found in the preceding MCMC solution of \citet{2023_10_Lissauer}. In order to better model the correlated noise present within each transit event, we may leverage out-of-transit observations, which locally share noise characteristics with those in-transit. Naturally, the question then arises as to how much out-of-transit data should be included when defining the phased photometric data window of any given fit? While there exists an abundance of available out-of-transit data, two constraining factors must be considered: (1) computational cost and (2) information gain. While only so much can be done in terms of computing power and software optimization, we can more deeply consider the notion that correlation between in-transit and out-of-transit noise decreases with increasing distance from the transit midpoint. To simplify things, the upper-bound on the \GP{}'s \lc{} prior can be set as the width of the phased photometric data window. It follows that a suitable choice of phased photometric data window width then preserves the \GP{}'s ability to accurately model timescales relevant to the transit event(s) and subsequent statistical meaning of \TGP{}-\GP{} model comparison; in other words, the possibility of solutions preferring longer-timescale fluctuations with little in-transit information is mitigated. 
    
    Core to the Bayesian approach of statistical model comparison is the consideration of a priori knowledge, as seen in \autoref{eq:Evidence Integral}. While intended to be used with informative (non-uniform) priors, it is not uncommon to lack the independent parameter constraints and distributions necessary for this. Such is the case in this study for all transit and noise model parameters, excluding \qone{} and \qtwo{}, whose prior distributions are precalculated with well-defined support according to \citet{2013_11_Kipping}. Varying searchable parameter space does not pose much of an issue apart from sampling inefficiencies when employing informative priors, as exploration beyond their regions of significant probability density with naturally well-defined support returns little to no information. The same cannot be said for uniform priors, whose normalization biases measurements of the \logB{} via \autoref{eq:Evidence Integral} \citep[for a detailed study regarding priors and caveats such as this within the context of Bayesian inference and model selection, see][]{2022_06_Llorente}. 

    To explore this implication, we propose a two-step method of quantile standardization to assess the potential bias induced by choices of \sigmap{} with fixed $\sigmad{} = 2$. First, overly-wide priors are cast using $\sigmap{} = 25$ in an attempt to fully capture the desired empirical posteriors. Once recovered, these can be used to provide accurate quantiles as: 
    \begin{equation} \label{eq:Sigma Quantile}
        q \left( \sigmap{} \right) \equiv \frac{1}{2} \left[ 1 + \rm{erf} \left( \frac{\sigmap{}}{\sqrt{2}} \right) \right] \times 100 \% \ ,
    \end{equation}
    with which the prior widths of subsequent runs can be defined; this notation is not to be confused with the limb-darkening parameters, \qone{} and \qtwo{}. In our code, these quantiles are set by the \corner{} \quantile{}\texttt{(x, q, weights)} function, with \texttt{x} and \texttt{weights} arguments given by $\sigmad{} = 2$, $\sigmap{} = 25$ \UltraNest{} empirical posteriors and \texttt{q} by $1 - q\left( \sigmap{} \right)$ or $q\left( \sigmap{} \right)$ for lower and upper bounds, respectively. A detailed example of this process with accompanying visuals can be found in \autoref{fig:Capturing Quantiles}.

    \begin{figure}
        \centering
        \includegraphics[height = \vsize]{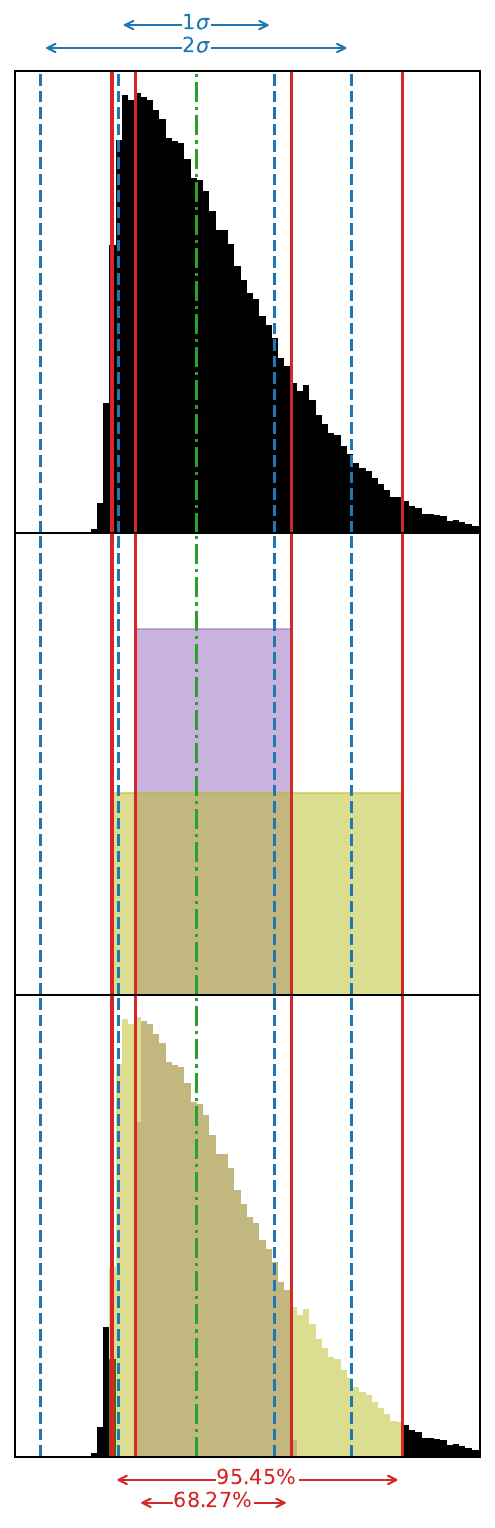}
    \end{figure}
    \begin{figure}[ht!]
        \centering
        \captionof{figure}[Capturing Quantiles]{A single-parameter mock example of the process used to obtain quantile-based prior widths for standardization tests conducted in \autoref{fig:Prior Width Variance Results}. In the first nested sampling run, a wide net is cast to ensure the parameter's empirical posterior distribution (black) is captured in its entirety (top). In subsequent runs, quantiles (solid red) of this complete posterior are used to define subsequent prior widths (middle), with which associated fractions of the complete posterior are recovered (bottom). While demonstrated with uniform priors, this is not a constraining factor; we apply this methodology to both uniform and non-uniform priors. In this example, skewed, non-Gaussian posterior behavior causes standard deviations (dashed blue) from the complete posterior's median (dashed-dotted green) to be inefficient in capturing parameter space information compared to quantiles. Since recovered posteriors are not guaranteed to be Gaussian in nature, we ensure consistent information gain by focusing only on relevant regions of parameter space via quantiles. The middle panel also visualizes how the normalized amplitude of a uniform prior is dictated by its width; this is a known origin of bias in \autoref{eq:Evidence Integral}.}
        \label{fig:Capturing Quantiles}
    \end{figure}

\begin{figure}[htb!]
        \centering
        \includegraphics[width = \hsize]{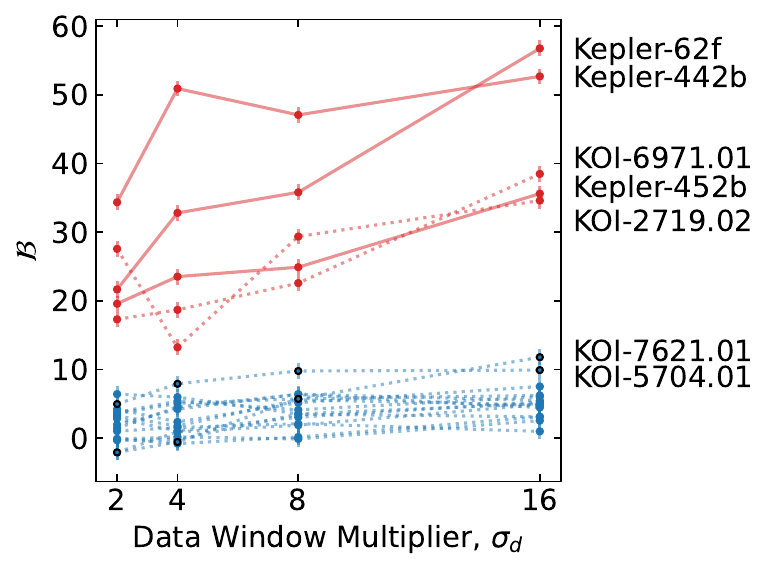}
        \caption[Data Window Variance Results]{The logged Bayes' factor, \logB{}, of each \KOI{} listed in \autoref{tab:Key Parameters Summary} as they vary with the phased photometric data window multiplier, \sigmad{}. Red and blue sub-populations correspond to strong \PC{}s and periodic transit events possessing inconclusive/weak evidence-based model preference with respect to the \PC{} and \FA{} hypotheses, respectively; solid lines indicate previously validated planets.}
        \label{fig:Data Window Variance Results}
    \end{figure}

    To assess correlation between the \logB{} and \sigmad{}, the sample population of \KOI{}s was fit for $\sigmad{} \in \left[ 2 , 4 , 8 , 16 \right]$ and fixed $\sigmap{} = 5$. The resulting emergence of two \KOI{} sub-populations was immediately apparent (see \autoref{fig:Data Window Variance Results}), these being strong \PC{}s with promising follow-up potential (red) and inconclusive/weak \PC{}s and/or \FA{}s (blue). Generally, the \PC{} group demonstrates positive trajectory with respect to \sigmad{} across all metrics whereas the latter group evolves in a relatively flat fashion. Here, the \logB{} exhibits a clearly defined sub-population boundary at a value of approximately 10. Overall, this behavior suggests that the dispositioning of \PC{}s from \FA{}s is largely independent of the chosen \sigmad{}, meaning that smaller phased photometric data windows may be favored to reduce the computational burden while retaining sufficient information; this additionally supports a transit-like timescale upper-bound on the \GP{}'s \lc{}. 

\begin{figure}[htb!]
        \centering
        \includegraphics[width = \hsize]{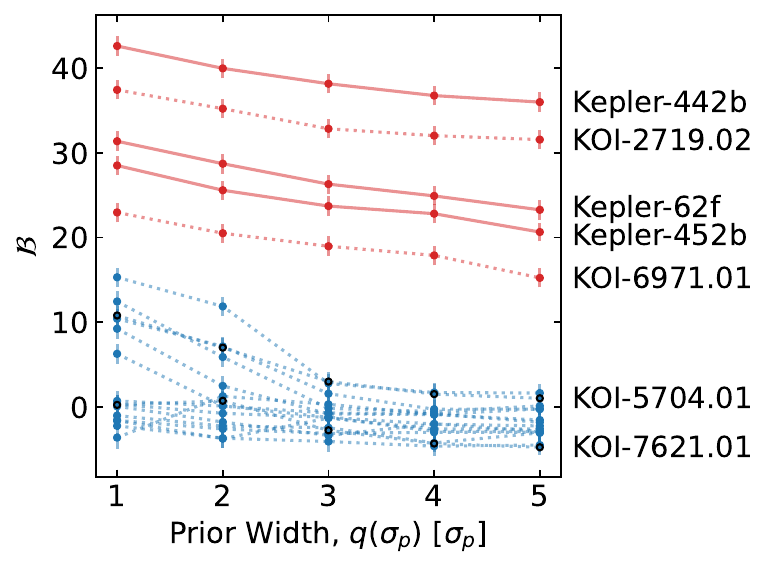}
        \caption[Prior Width Variance Results]{Same as \autoref{fig:Data Window Variance Results}, but with respect to \sigmap{}, which sets quantile-based prior widths.}
        \label{fig:Prior Width Variance Results}
    \end{figure}

    In testing how the variance of prior width affects our results, we use quantiles of $\sim$ $68.2689 \%$, $95.4410 \%$, $99.7300 \%$, $99.9937 \%$, and $99.9999 \%$ for prior widths of subsequent independently computed nested sampling runs corresponding to $\sigmap{} \in \left[ 1 , 5 \right]$ and fixed $\sigmad{} = 2$. As expected, the \logB{} possesses a general inverse proportionality to \sigmap{} resulting from \autoref{eq:Evidence Integral} (see \autoref{fig:Prior Width Variance Results}). Conveniently however, this trend is roughly similar across the entire \KOI{} sample population, indicating that sample populations with prior widths set by a consistent/shared choice of quantile will experience a population-wide shift in \logB{} plus some small variance of $\sim \pm 5$ such that relative ranking between targets are still valid. Interestingly, it seems that this variance is reduced to stochastic order with greater probability of \PC{} status, meaning that we can be relatively more confident in our findings for the red sub-population. While these tests must be performed on a much larger \KOI{} population to draw any statistically robust conclusions, if the general inverse trend is to hold, one could theoretically fit and recover an empirical correction relating the \logB{} and \sigmap{}. For now, there does not seem to be a preferred \sigmap{} in terms of bias reduction, but we can recommend quantiles corresponding to at least $\sigmap{} = 3$ in order to promote sufficient prior-posterior information gain and properly recovered posteriors.

    Finally, \autoref{fig:Data Window Variance Results} and \autoref{fig:Prior Width Variance Results} both suggest the presence of five strong \PC{}s: three known exoplanets, Kepler-62f, Kepler-442b, and Kepler-452b, and two new additions residing within the habitable zones of their host stars, \KOI{}-2719.02 and \KOI{}-6971.01. \KOI{}-5704.01 and \KOI{}-7621.01 are also noteworthy: the former consistently crests the \logB{} boundary of $\sim 10$ whereas the latter displays anomalous behavior in that it draws a steep upward trajectory in \logB{} characteristic of the strong \PC{} sub-population across \sigmad{} while only escaping the otherwise clearly defined boundary at $\sigmad{} = 16$; this is not observed for any other member of the inconclusive/weak sub-population. 
    
    If verified with current fitted parameters (see \autoref{tab:Key Parameters Summary}), \KOI{}-7621.01 would rank alongside Kepler-62f, Kepler-442b, and Kepler-452b in terms of Earth--Sun analog candidacy. That being said, some points of contention must be addressed regarding this candidate: (1) we derive a nearly parabolic minimum eccentricity of $\eccentricity{}_{\rm min} = 0.89_{- 0.04}^{+ 0.02}$ and (2) the photometric data contains long-timescale fluctuations of considerable amplitude, likely caused by spot modulation. Since eccentricity is degenerate with the mean stellar density and impact parameter---found to be $\rhostar{} = 76_{- 31}^{+ 26} \ \gcmthree{}$ and $\bimpact{} = 0.33_{- 0.23}^{+ 0.28}$, whereas \citet{2020_06_Berger} obtained $\rhostar{} = 1.00 \pm 0.11 \ \gcmthree{}$---, we cannot readily conclude whether this is a highly-eccentric orbit or the consequence of grazing transits. 
    
    In terms of obtaining best-fitting model parameters, fluctuations caused by spot modulation affect a greater number of data points than those caused by transit-timescale events such that our \GP{}'s \lc{} is motivated toward longer-timescales, thereby foregoing the ability to represent transit events in favor of producing an overall ``better'' fit. It follows in subsequent \TGP{}-\GP{} model comparison that the \TGP{} model will always outperform the \GP{}-only solution. We can then expect the \logB{} to become artificially inflated with increasing \sigmad{} in the presence of long-timescale correlated noise. Of the strong \PC{} sub-population, \KOI{}-2719.02 is the only target to experience this spot modulation inflation effect; its strong \PC{} status is not invalidated however, as even with a sub-unity \GP{} \lc{} from the $\sigmad{} = 2$ and $\sigmap{} = 5$ solution, its \logB{} remains significant in magnitude. 
    
    Our future work will adopt more aggressive data preconditioning techniques (e.g., application of low-frequency bandpass filters) in an effort to mitigate \logB{} inflation by focusing the \GP{} on transit-like timescales. All things considered, both \KOI{}-5704.01 and \KOI{}-7621.01 certainly warrant further investigation in future studies.

\subsection[Comparisons to Multiple Event Statistic and Signal-to-Noise Ratio]{Comparisons to Multiple Event Statistic and Signal-to-Noise Ratio} \label{subsec:Comparisons to Multiple Event Statistic and Signal-to-Noise Ratio}

\begin{figure*}[htb!]
        \centering
        \includegraphics[width = \hsize]{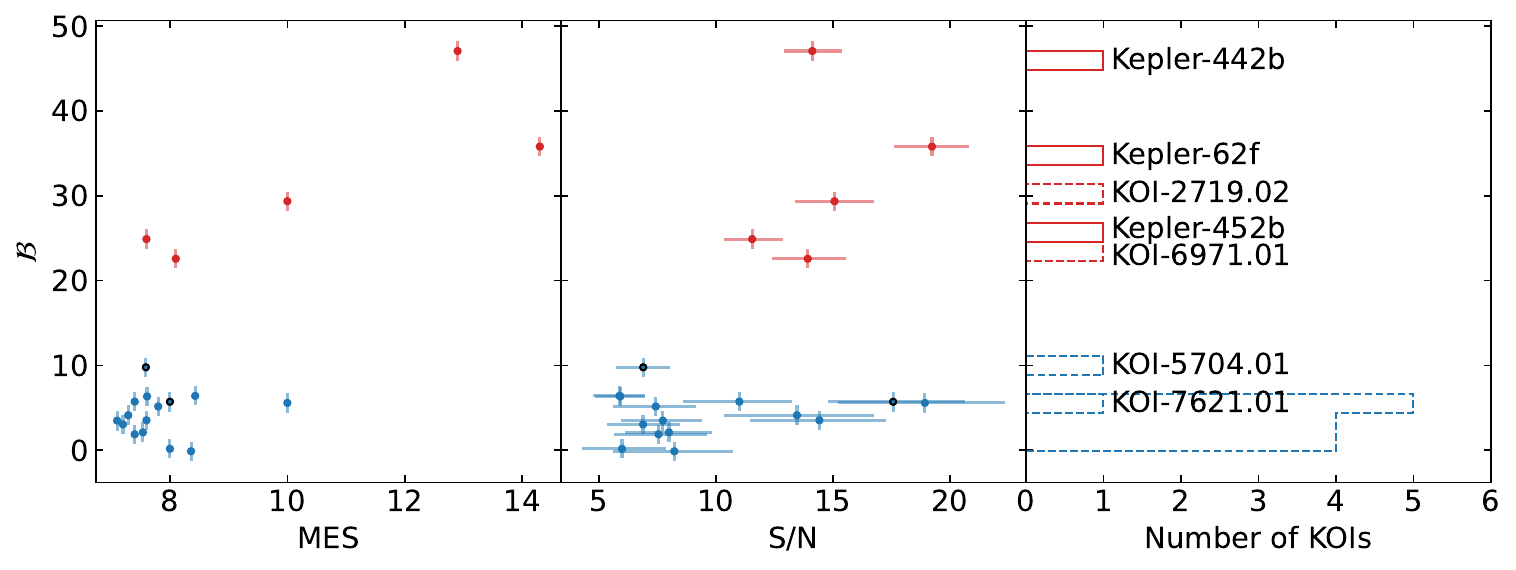}
        \caption[Evidence, MES, and S/N]{DR25 \MES{} (left) and our \SNR{} (middle) metrics compared against our logged Bayes's factor, \logB{}, and their histogram (right) for the \KOI{} sample given in \autoref{tab:Key Parameters Summary}. Red and blue sub-populations correspond to strong \PC{}s and periodic transit events possessing inconclusive/weak evidence-based model preference with respect to the \PC{} and/or \FA{} hypotheses, respectively; solid lines indicate previously validated planets. Note the ample separation between strong \PC{}s and inconclusive/weak \PC{}s (or \FA{}s) revealed by the \logB{}, which \MES{} and \SNR{} are otherwise blind to.}
        \label{fig:MES-SNR-logB}
    \end{figure*}

    For our final result, we empirically compare the Bayesian evidence approach against Q1~--~16 (or DR25 where available) \MES{} and our derived \SNR{} according to Equation~5 of \citet{2015_03_Rowe}; the latter two being standard metrics for candidate discrimination in \TCE{}/\KOI{} searches and catalogs. \autoref{fig:MES-SNR-logB} shows that our novel methodology presents the ability to clearly distinguish between strong \PC{}s (red) and periodic transit events possessing inconclusive/weak evidence-based model preference with respect to the \PC{} and \FA{} hypotheses (blue) in regimes contaminated by high-levels of white and correlated noise, as demonstrated by the ample separation of these two sub-populations. This is in contrast to both the \MES{} and \SNR{}, which completely mix populations, save for Kepler-62f and Kepler-442b as the only \KOI{}s that we analyzed with $\MES{} > 10$. Although we observe no strong \PC{}s with derived $\SNR{} < 10$, the appearance of inconclusive/weak \PC{}s or \FA{}s beyond this threshold illustrates a potential deficiency of \SNR{} when used as a \PC{}-\FA{} discrimination metric in comparison to the \logB{}. In the context of choosing \MES{} or \SNR{} cutoffs for searches, this means that strong \PC{}s are likely to be lost and/or inconclusive/weak \PC{}s and/or \FA{}s included. Should this clear separation between populations holds across larger \KOI{} samples, the \logB{} could substantially reduce this blind spot.

\section[Conclusions]{Conclusions} \label{sec:Conclusions}

    Our analysis of targets via the simultaneous modeling of transits alongside a combined white and correlated noise \GP{} yields fundamental transit parameters (e.g., scaled planetary radius, \RpRstar{}) and Bayesian evidence-driven \PC{}-\FA{} model comparison in the most robust approach to date. It is then important to note that there is a discrepancy between our results and those of DR25, as illustrated by the case of Kepler-452b. While DR25 yields an estimated reliability value of $\sim 40 \%$ \citep[see Figure~11 of][]{2018_04_Thompson}, we recover a strongly favored \PC{} status of $\logB{} = 24.9_{- 1.2}^{+ 1.1}$. \textit{Note that the Bayesian evidence does not directly translate to a reliability percentage, so here we simply compared the interpretations of their results}; however, we aim to create a mapping between the two in future works. To do so, we next plan to conduct extensive injection-recovery testing, which will also facilitate an understanding of the \logB{} floor from strong \FA{}s.
    
    Having performed Bayesian model comparison between \PC{} and \FA{} hypotheses on the periodic transit-like photometric events of each \KOI{} listed in \autoref{tab:Key Parameters Summary}, we report strong \PC{} dispositions of Kepler-62f, Kepler-442b, and Kepler-452b---agreeing with preexisting studies---plus the two new additions of \KOI{}-2719.02 and \KOI{}-6971.01, as well as two moderately strong \PC{}s, \KOI{}-5704.01 and \KOI{}-7621.01.

    Preliminary testing indicates a demand for the choice of free model parameters, \sigmad{} and \sigmap{}, to be shared across any given sample population of \KOI{}s in order to promote statistically sound comparisons between targets. Furthermore, smaller phased photometric data windows (lower \sigmad{}) and consistent quantile-based prior widths likely mitigate potential biases. 

    The recovered posteriors of fitted/derived parameters were used to obtain a statistical description of the \SNR{} with uncertainties on a per-target basis, rather than its point-estimate counterpart commonly reported in previous studies. The \TGP{} approach also yields similar to significantly improved values of \SNR{} with respect to those reported by \citet{2023_10_Lissauer} (e.g., see \KOI{}-2719.02 and \KOI{}-7621.01 in \autoref{tab:Key Parameters Summary}).

    That being said, both \MES{} and \SNR{} exhibited vulnerability to candidate misidentification, whereas the \logB{} was able to clearly distinguish strong \PC{}s from inconclusive/weak \PC{}s and/or \FA{}s. Regardless of whether the \logB{} is adopted as a standard metric in \PC{}-\FA{} dispositioning, the \MES{} and \SNR{} should undergo additional investigation and be used thoughtfully. 

\section[Next Steps]{Next Steps} \label{sec:Next Steps}

    When allocated one node (32 CPU threads) on high-performance computing clusters, our current nested sampling infrastructure sees typical per-target timescales on the order of a week. As such, our future work will instead rely on the development of a Simulation-Based Inference \citep[\SBI{};][]{2020_12_Cranmer} machine learning infrastructure; these have seen great success in recent years \citep[see][]{2018_07_Alsing,2019_09_Alsing,2020_08_Tejero_Cantero,2020_11_Miller,2022_07_Miller,2023_01_Legin}. The amortized nature of \SBI{} will allow for computationally efficient deployment across parameter space in catalog-wide applications to current and future missions (\ikt, \Ktwot, \tesst, \textit{PLATO}, etc.). Cutting-edge supporting frameworks/methodologies \citep[see][]{2021_11_McEwen,2023_05_Jeffrey,2023_06_Legin} will facilitate the core Bayesian evidence-based approach debuted here. 

\newpage

\section*{Acknowledgments} \label{sec:Acknowledgments}

    We thank \href{https://orcid.org/0000-0003-0081-1797}{Steve Bryson\hskip2pt\includegraphics[width=9pt]{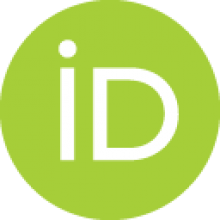}} and \href{https://orcid.org/0000-0003-3544-3939}{Laurence Perreault-Levasseur\hskip2pt\includegraphics[width=9pt]{Orcid-ID.png}} for their assistance, consultation, and contributions. MRBM received financial support from a Bishop's University Foundation Graduate Entrance Scholarship and a Fondation Arbour Master's Scholarship.  JFR acknowledges support from the NSERC Discovery program and the Canada Research Chair program. This research made use of \DRAC{} computer resources from a RAC allocation to JFR. This research has made use of the \NASAExoplanetArchive{}, which is operated by the California Institute of Technology under contract with the National Aeronautics and Space Administration through the Exoplanet Exploration Program. The specific observations analyzed can be accessed via \dataset[DOI: 10.26093/cds/vizier.22350038]{https://doi.org/10.26093/cds/vizier.22350038} \citep{2018_04_Thompson,2018_08_Thompson}.

\facilities{
    \NASAADS{},
    \KeplerSpaceTelescope{} \citep{2010_02_Borucki,2010_04_Koch,2016_03_Borucki},
    \NASAExoplanetArchive{} \citep{2022_11_NASA_Exoplanet_Archive}
}

\software{
    \ChainConsumer{} \citep{2016_08_Hinton};
    \corner{} \citep{2016_06_Foreman_Mackey};
    \Matplotlib{} \citep{2007_05_Hunter};
    \Numba{} \citep{2015_11_Lam};
    \NumPy{} \citep{2020_09_Harris};
    \pandas{} \citep{2010_McKinney,2020_02_The_Pandas_Development_Team};
    \Python{} \citep{1995_01_a_Van_Rossum,1995_01_b_Van_Rossum,1995_01_c_Van_Rossum,1995_01_d_Van_Rossum,1996_05_Dubois,2007_01_Oliphant};
    \SciPy{} \citep{2020_02_Virtanen};
    \transitfitfive{} \citep{2016_08_Rowe};
    \UltraNest{} \citep{2016_01_Buchner,2019_11_Buchner,2021_04_Buchner}
}

\section*{Appendix A \\ Terminology and Acronyms} \label{sec:Terminology and Acronyms}

    A list of terminology and acronyms alongside their corresponding definitions can be found here.

    \begin{enumerate}
        \item \hypertarget{def:wperi}{\textbf{Argument of Periapsis}, \wperi{}: Angle from ascending node to periapsis along the direction of motion for a given orbiting body.}
        \item \hypertarget{def:evidence}{\textbf{Bayesian Evidence}, $\boldsymbol{\evidence{}}$: The model-agnostic probability of observing the data, \data{} (see \autoref{eq:Bayes's Theorem} and \autoref{eq:Evidence Integral}).}
        \item \hypertarget{def:tTc}{\textbf{Central Transit Duration}, $\boldsymbol{\tTc{}}$: The duration, \tT{}, of an equatorial $\left( \bimpact{} = 0 \right)$ transit.}
        \item \hypertarget{def:sigmac}{\textbf{Correlated Noise Amplitude Scale}, $\boldsymbol{\sigmac{}}$: The Mat{\'e}rn 3/2 kernel \GP{} unitless amplitude scale (see \autoref{eq:Matern 3/2}; fitted noise model parameter).} 
        \item \hypertarget{def:lc}{\textbf{Correlated Noise Length Scale}, $\boldsymbol{\lc{}}$: The Mat{\'e}rn 3/2 kernel \GP{} unitless length scale (see \autoref{eq:Matern 3/2}; fitted noise model parameter).} 
        \item \hypertarget{def:sigmad}{\textbf{Data Window Width Free Parameter}, $\boldsymbol{\sigmad{}}$: The factor used to define the phased photometric data window in terms of transit durations, \tT{}'s, out from the transit midpoint as discussed in \autoref{subsec:Varying Free Parameters}.}
        \item \hypertarget{def:eccentricity}{\textbf{Eccentricity}, $\boldsymbol{\eccentricity{}}$: The value describing orbital shape, ranging from circular $\left( \eccentricity{} = 0 \right)$ to elliptical $\left( 0 < \eccentricity{} < 1 \right)$ to parabolic $\left( \eccentricity{} = 1 \right)$.}
        \item \hypertarget{def:ecosw}{\hypertarget{def:esinw}{\textbf{Eccentricity Projections}, $\boldsymbol{\ecosw{}}$ and $\boldsymbol{\esinw{}}$: Transit model parameters for non-circular orbits representing eccentricity, \eccentricity{}, vector components projected with respect to the argument of periapsis, \wperi{}.}}
        \item \hypertarget{def:Teff}{\textbf{Effective Temperature}, $\boldsymbol{\Teff{}}$: The average surface temperature of an object (e.g., star), given in this study by \citet{2020_06_Berger}.}
        \item \hypertarget{def:Tzero}{\textbf{Ephemeris}, $\boldsymbol{\Tzero{}}$: The \KOI{}'s time series epicenter in units of \days{} (fitted transit model parameter).} 
        \item \hypertarget{def:etaEarth}{\hypertarget{def:etaVenus}{\textbf{Eta-Earth} (or \textbf{Eta-Venus}), $\boldsymbol{\etaEarth{}} \left( \text{or } \boldsymbol{\etaVenus{}} \right)$: The occurrence rate of Earth-like (or Venus-like) planets around Sun-like stars.}}
        \item \hypertarget{def:FA}{\textbf{False-Alarm}, $\boldsymbol{\FA{}}$: A periodic transit-like signal caused by instrumental and/or stellar noise/variability.} 
        \item \hypertarget{def:FP}{\textbf{False-Positive}, $\boldsymbol{\FP{}}$: A periodic transit-like signal caused by physical sources other than a transiting exoplanet (e.g., eclipsing stellar binary signatures).} 
        \item \hypertarget{def:GP}{\textbf{Gaussian process Model}, $\boldsymbol{\GP{}}$: The \FA{} hypothesis model (see \autoref{tab:Model Parameter Descriptions}).}
        \item \hypertarget{def:bimpact}{\textbf{Impact Parameter}, $\boldsymbol{\bimpact{}}$: The unitless projection of \aRstar{} with respect to the orbital inclination, $i$ (fitted transit model parameter).}
        \item \hypertarget{def:Szero}{\textbf{Insolation Flux}, $\boldsymbol{\Szero{}}$: The measure of incident solar radiation on a surface or body (e.g., exoplanet; see \autoref{eq:Insolation Flux}).}
        \item \hypertarget{def:KOI}{\textbf{Kepler Object of Interest}, $\boldsymbol{\KOI{}}$: A periodic transit-like event or \TCE{} that warrants further review \citep{2018_04_Thompson}.} 
        \item \hypertarget{def:likelihood}{\textbf{Likelihood}, $\boldsymbol{\likelihood{}}$: The probability which quantifies how strongly the data, \data{}, supports the modelled parameters, \parameters{} (see \autoref{eq:Bayes's Theorem}).} 
        \item \hypertarget{def:qone}{\hypertarget{def:qtwo}{\textbf{Limb-darkening}, $\boldsymbol{\qone{}}$ and $\boldsymbol{\qtwo{}}$: The unitless \citet{2013_11_Kipping} reparameterization of \citet{2002_12_Mandel} limb-darkening (fitted transit model parameters).}}
        \item \hypertarget{def:logB}{\textbf{Logged Bayes's Factor}, $\boldsymbol{\logB{}}$: The logged Bayes's factor representing the difference in logged Bayesian evidences, \evidence{}, between any two models (e.g., \TGP{} and \GP{}) applied to the same data set, \data{} (see \autoref{eq:Bayes's Factor}, \autoref{eq:Logged Bayes's Factor} and \autoref{eq:logB}).}
        \item \hypertarget{def:MAP}{\textbf{Maximum a Posteriori}, $\boldsymbol{\MAP{}}$: The most probable set of modelled parameters as given by Bayes's theorem (see \autoref{eq:Bayes's Theorem}).}
        \item \hypertarget{def:rhostar}{\textbf{Mean Stellar Density}, $\boldsymbol{\rhostar{}}$: The mean stellar density in units of \gcmthree{} (fitted transit model parameter).}
        \item \hypertarget{def:MES}{\textbf{Multiple Event Statistic}, $\boldsymbol{\MES{}}$: A measure describing the combined significance of all observed transits in the detrended and whitened light curve with the assumption of a linear ephemeris, \Tzero{} \citep{2002_08_Jenkins,2018_04_Thompson}.}
        \item \hypertarget{def:Period}{\textbf{Orbital Period}, $\boldsymbol{\Period{}}$: The \KOI{}'s time series orbital period in units of \days{} (fitted transit model parameter).} 
        \item \hypertarget{def:Fzero}{\textbf{Photometric Zero-Point}, $\boldsymbol{\Fzero{}}$: The relative (unitless) photometric zero-point offset (fitted transit model parameter).} 
        \item \hypertarget{def:PC}{\textbf{Planet-Candidate}, $\boldsymbol{\PC{}}$: A \KOI{} which has passed \FA{} vetting procedures but has yet to undergo/pass \FP{} vetting and/or be confirmed by alternative observation techniques.} 
        \item \hypertarget{def:Rp}{\textbf{Planetary Radius}, $\boldsymbol{\Rp{}}$: The radius of the companion exoplanet.}
        \item \hypertarget{def:posteriors}{\textbf{Posterior Distribution}, $\boldsymbol{\posteriors{}}$: The updated probability of modelled parameters, \parameters{}, given new data, \data{}, and informed by combining the likelihood, \likelihood{}, priors, \priors{}, and Bayesian evidence, \evidence{} (see \autoref{eq:Bayes's Theorem}).} 
        \item \hypertarget{def:priors}{\textbf{Prior Distribution}, $\boldsymbol{\priors{}}$: The initial probability or belief about given model parameters, \parameters{}, before any new data, \data{}, is taken into account (see \autoref{eq:Bayes's Theorem}).}  
        \item \hypertarget{def:sigmap}{\textbf{Prior Width Free Parameter}, $\boldsymbol{\sigmap{}}$: The factor used to define fitted parameter prior widths for \UltraNest{} in terms of MCMC-recovered standard deviations with respect to their maximum-likelihood-estimator values as given by \citet{2023_10_Lissauer} before quantile-defined widths are obtained using \autoref{eq:Sigma Quantile} as discussed in \autoref{subsec:Varying Free Parameters}.}
        \item \hypertarget{def:RpRstar}{\textbf{Scaled Planetary Radius}, $\boldsymbol{\RpRstar{}}$: The unitless ratio of companion planetary and host stellar radii (fitted transit model parameter).}   
        \item \hypertarget{def:aRstar}{\textbf{Scaled Semi-Major Axis}, $\boldsymbol{\aRstar{}}$: Unitless ratio of the companion exoplanet's semi-major axis scaled with respect to the host's stellar radius.}
        \item \hypertarget{def:SNR}{\textbf{Signal-to-Noise}, $\boldsymbol{\SNR{}}$: The quantification of a desired signal's quality with respect to the level of unwanted noise contamination \citep[see Equation~5 of][]{2015_03_Rowe}.}
        \item \hypertarget{def:Rstar}{\textbf{Stellar Radius}, $\boldsymbol{\Rstar{}}$: The radius of the host star, given in this study by \citet{2020_06_Berger}.}
        \item \hypertarget{def:TCE}{\textbf{Threshold Crossing Event}, $\boldsymbol{\TCE{}}$: A periodic signal identified by the Transiting Planet Search \citep[TPS;][]{2010_07_Jenkins,2016_12_Twicken,2020_03_Jenkins} module of the Kepler Science Operations Center (SOC) Science Processing Pipeline \citep{2010_04_Jenkins}.}
        \item \hypertarget{def:TGP}{\textbf{Transit plus Gaussian process Model}, $\boldsymbol{\TGP{}}$: The \PC{} hypothesis model (see \autoref{tab:Model Parameter Descriptions}).}
        \item \hypertarget{def:tT}{\textbf{Transit Duration}, $\boldsymbol{\tT{}}$: The total time taken for the (exoplanet) companion to occult its host (star) from ingress to egress (i.e., beginning to end; see \autoref{eq:Transit Duration rho}).}
        \item \hypertarget{def:sigmaw}{\textbf{White Noise Amplitude Scale}, $\boldsymbol{\sigmaw{}}$: The unitless scaling factor to DR25 reported photometric errors (fitted noise model parameter).} 
    \end{enumerate}

\section*{Appendix B \\ Strong and Weak Cases Supplementary Figures} \label{sec:Strong and Weak Cases Figures}

\begin{figure*}[htb!]
        \centering
        \includegraphics[width = \hsize]{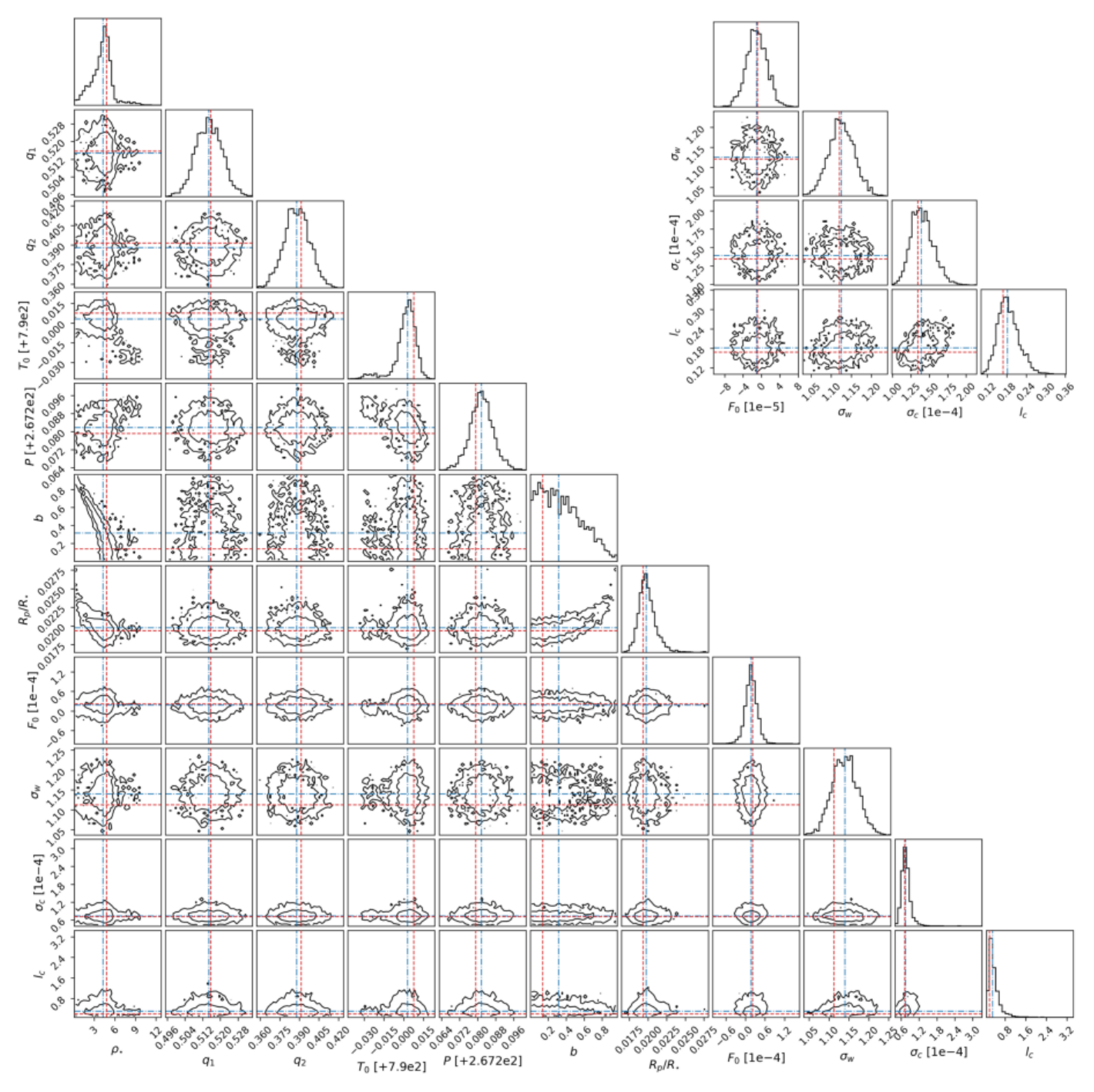}
        \caption[Kepler-62f Corner Plots]{$\sigmad{} = 8$, $\sigmap{} = 5$ Kepler-62f \TGP{} (bottom-left) and \GP{} (top-right) corner plots depicting parameter behavior/covariance, with empirical posterior distributions overlaid by \MAP{} (red) and median (blue) solutions above each column.}
        \label{fig:Kepler-62f Corner Plots}
    \end{figure*}

\begin{figure*}[htb!]
        \centering
        \includegraphics[width = \hsize]{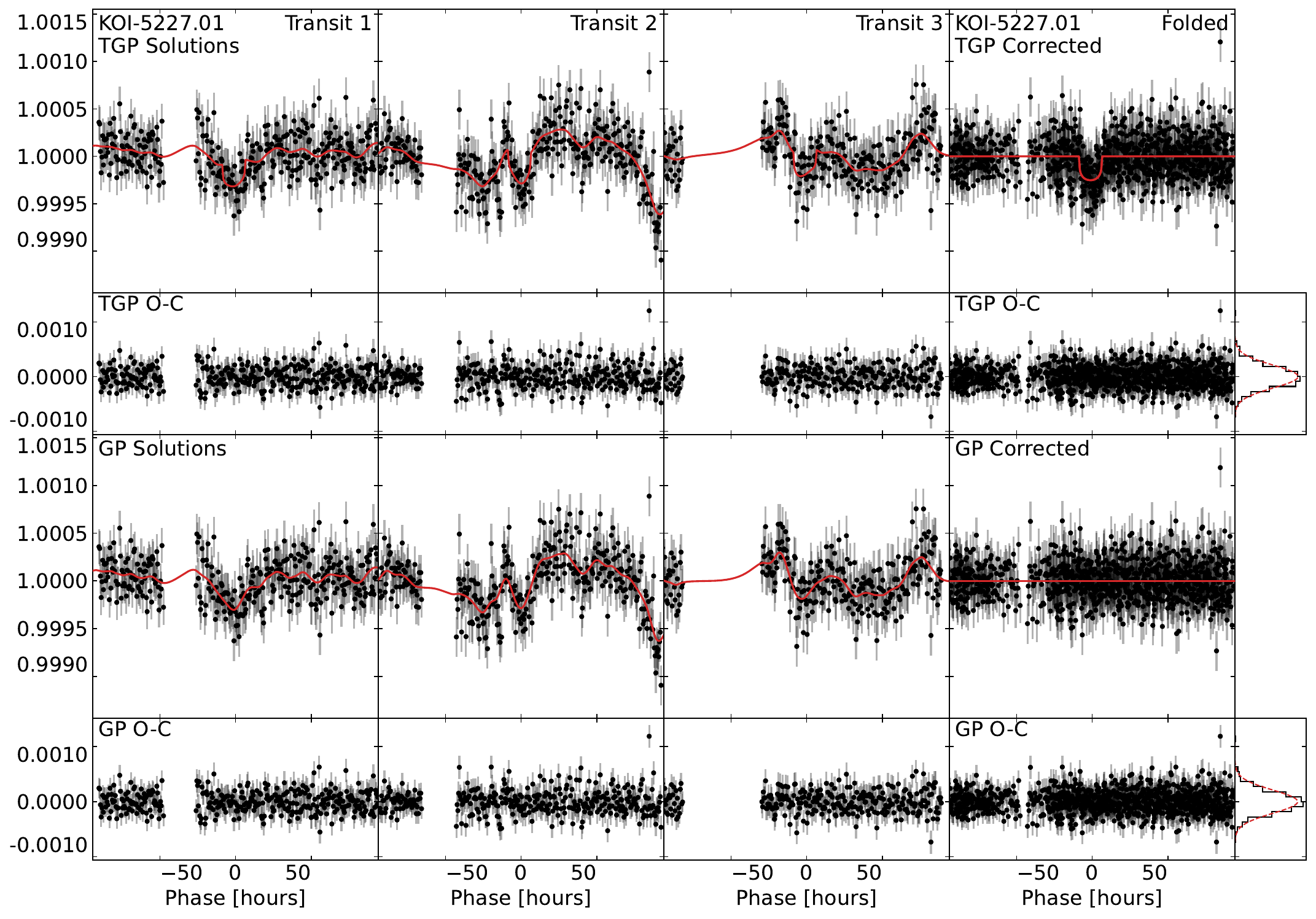}
        \caption[KOI-5227.01 Phase Plots]{Same as \autoref{fig:Kepler-62f Phase Plots}, but for \KOI{}-5227.01.}
        \label{fig:KOI-5227.01 Phase Plots}
    \end{figure*}

\begin{figure*}[htb!]
        \centering
        \includegraphics[width = \hsize]{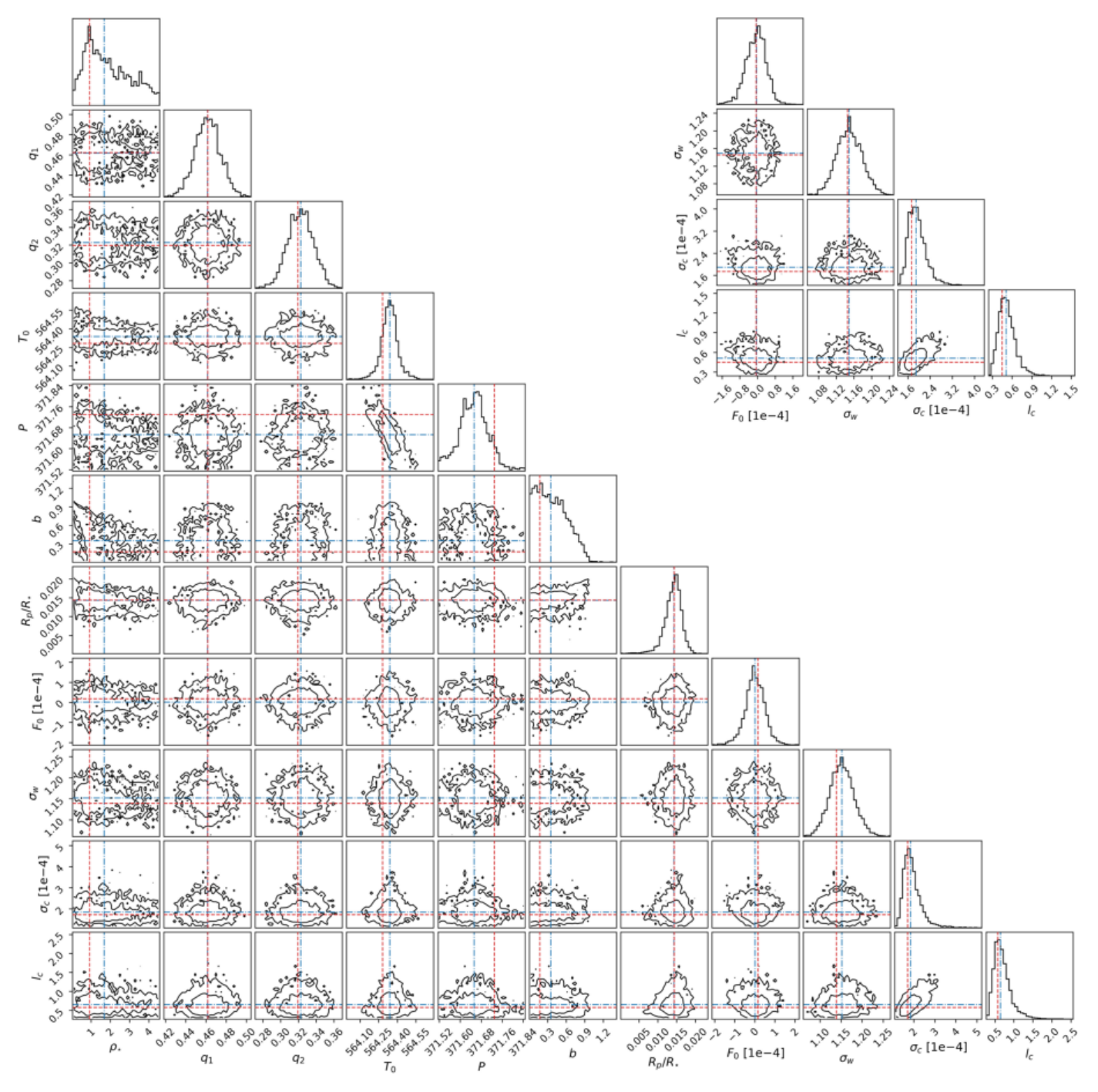}
        \caption[KOI-5227.01 Corner Plots]{Same as \autoref{fig:Kepler-62f Corner Plots}, but for \KOI{}-5227.01.}
        \label{fig:KOI-5227.01 Corner Plots}
    \end{figure*}

    Appendix B contains \autoref{fig:Kepler-62f Corner Plots} -- \autoref{fig:KOI-5227.01 Corner Plots}.

    \FloatBarrier

    \FloatBarrier

\section*{ORCID iDs}

    \footnotesize{\noindent Michael R. B. Matesic \includegraphics[width=9pt]{Orcid-ID.png} \href{https://orcid.org/0000-0002-1119-7473}{https://orcid.org/0000-0002-1119-7473} \\
    Jason F. Rowe \includegraphics[width=9pt]{Orcid-ID.png} \href{https://orcid.org/0000-0002-5904-1865}{https://orcid.org/0000-0002-5904-1865} \\
    John H. Livingston \includegraphics[width=9pt]{Orcid-ID.png} \href{https://orcid.org/0000-0002-4881-3620}{https://orcid.org/0000-0002-4881-3620} \\
    Shishir Dholakia \includegraphics[width=9pt]{Orcid-ID.png} \href{https://orcid.org/0000-0001-6263-4437}{https://orcid.org/0000-0001-6263-4437} \\
    Daniel Jontof-Hutter \includegraphics[width=9pt]{Orcid-ID.png} \href{https://orcid.org/0000-0002-6227-7510}{https://orcid.org/0000-0002-6227-7510} \\
    Jack J. Lissauer \includegraphics[width=9pt]{Orcid-ID.png} \href{https://orcid.org/0000-0001-6513-1659}{https://orcid.org/0000-0001-6513-1659}}

\bibliography{bibliography}{}
\bibliographystyle{aasjournal}

\clearpage

\end{document}